\documentclass[preprint]{elsarticle}

\usepackage{lineno,hyperref}
\modulolinenumbers[5]

\usepackage[nointegrals]{wasysym}

\usepackage{amsmath, amsthm, amssymb} 

\begin{document}

\begin{frontmatter}

\title{Localisation of gamma-ray interaction points in thick monolithic CeBr$_3$ and LaBr$_3$:Ce scintillators}

\author[ucd]{Alexei Ulyanov \corref{cor1}}
\cortext[cor1]{Corresponding author. E-mail address: alexey.uliyanov@ucd.ie}

\author[ucd,gmit]{Oran Morris}
\author[ucd]{Oliver J. Roberts} 
\author[ucd]{Isaac Tobin} 
\author[ucd]{Lorraine Hanlon}
\author[ucd]{Sheila McBreen} 
\author[ucd]{David Murphy} 
\author[esa]{Nick Nelms} 
\author[esa]{Brian Shortt} 

\address[ucd]{School of Physics, University College Dublin, Belfield, Dublin 4, Ireland}
\address[gmit]{Department of Computer Science \& Applied Physics, Galway-Mayo Institute of Technology, Galway, Ireland}
\address[esa]{European Space Agency, ESTEC, 2200 AG Noordwijk, The Netherlands}

\begin{abstract}
Localisation of gamma-ray interaction points in monolithic scintillator crystals can simplify the design and improve the performance of a future Compton telescope for gamma-ray astronomy. In this paper we compare the position resolution of three monolithic scintillators: a $28 \times 28 \times 20$~mm$^3$ (length $\times$ breadth $\times$ thickness) LaBr$_3$:Ce crystal, a $25 \times 25 \times 20$~mm$^3$  CeBr$_3$ crystal and a $25 \times 25 \times 10$~mm$^3$  CeBr$_3$ crystal. Each crystal was encapsulated and coupled to an array of $4\times4$ silicon photomultipliers through an optical window. The measurements were conducted using 81~keV and 356~keV gamma-rays from a collimated $^{133}$Ba source. The 3D position reconstruction of interaction points was performed using artificial neural networks trained with experimental data. Although the position resolution was significantly better for the thinner crystal, the 20~mm thick CeBr$_3$ crystal showed an acceptable resolution of about 5.4~mm FWHM for the $x$ and $y$ coordinates, and 7.8~mm FWHM for the $z$-coordinate (crystal depth) at 356~keV. These values were obtained from the full position scans of the crystal sides. The position resolution of the  LaBr$_3$:Ce crystal was found to be considerably worse, presumably due to the highly diffusive optical interface between the crystal and the optical window of the enclosure. The energy resolution (FWHM) measured for 662~keV gamma-rays was 4.0\%  for LaBr$_3$:Ce and 5.5\% for CeBr$_3$. The same crystals equipped with a PMT (Hamamatsu R6322-100) gave an energy resolution of 3.0\% and 4.7\%, respectively.

\end{abstract}

\begin{keyword}
\texttt{ position resolution \sep spatial resolution \sep gamma-ray \sep scintillator \sep cerium bromide \sep silicon photomultiplier \sep ANN} 
\end{keyword}

\end{frontmatter}

\clearpage
\noindent
\copyright 2016. The manuscript is licensed under the CC-BY-NC-ND 4.0 license \\http://creativecommons.org/licenses/by-nc-nd/4.0/

\vspace*{5cm}
\noindent
The published article is available at http://dx.doi.org/10.1016/j.nima.2016.11.025

\clearpage
\section{Introduction}

Due to the low interaction cross-sections and difficulties of reconstructing Compton events, the photon energy range of 0.2-50~MeV is currently the least explored band in high-energy astronomy. At the same time, this is a unique energy band that can provide answers to many fundamental questions regarding the synthesis of matter in the Universe, the mechanisms behind supernova explosions, the origin of positrons in our galaxy and the nature of radiation processes and particle acceleration in extreme cosmic sources including gamma-ray bursts, pulsars and magnetars. In order to address these questions, a number of new missions have been recently proposed to the European Space Agency (ESA), such as DUAL~\cite{dual}, CAPSITT~\cite{capsitt}, GRIPS~\cite{greiner} and ASTROGAM~\cite{astrogam}. The combined Compton and pair creation telescopes proposed for the GRIPS and ASTROGAM missions rely on a silicon strip tracker and a position sensitive calorimeter. For pair conversion events (above 10~MeV), the electron and positron tracks in the silicon detectors are used to reconstruct the initial gamma-ray direction. The total energy is measured together by the tracker and the calorimeter. For Compton events (0.2-20~MeV), the tracker identifies the interaction points of initial gamma-rays and measures the energy of recoil electrons. The Compton-scattered gamma-rays are detected by the calorimeter, but may also have additional interactions in the tracker. The energy resolution of the calorimeter is important for reconstruction of Compton events and affects both the spectral and angular resolution of the Compton telescope. Because of its excellent energy resolution, cerium doped lanthanum bromide (LaBr$_3$:Ce) was proposed as a scintillator material for the calorimeter in the GRIPS mission. However, due to a strong radiation background caused by the decays from the naturally occurring $^{138}$La isotope~\cite{quarati_radioactivity}, other scintillator materials with good energy resolution are now being considered. In particular, recent studies of cerium bromide crystals doped with strontium or calcium suggest that these new scintillators may provide an attractive alternative to LaBr$_3$:Ce~\cite{quarati_cebr3_codoped}.

In addition to good energy resolution, the calorimeter is required to measure the interaction points of scattered gamma-rays with an accuracy of about 1~cm~\cite{ulyanov_acta2013}. The required position resolution can be achieved using sufficiently small scintillator crystals, or alternatively, using large monolithic crystals coupled to multi-pixel photodetectors~\cite{gostojic2013}. In the latter case, the 3D position of a gamma-ray interaction point in a crystal can be reconstructed from the spatial distribution of the scintillation light across the photodetector. This method is known to work very well for relatively thin scintillators  (10~mm or less)~\cite{bruy2003, pani2006, seifert2012, cabello2013} and is actively employed in new detectors that are being developed for medical imaging applications~\cite{thirolf2014, llosa2016, borghi2016}. 
Efficient detection of high-energy gamma-rays requires thicker calorimeters. Recently, a position resolution of about 1.5~mm FWHM at 511~keV has been demonstrated for a 22~mm thick LYSO:Ce crystal coupled to a digital silicon photomultiplier array~\cite{borghi2016}. LYSO:Ce crystals, however, are not well suited for Compton telescopes because of the inferior energy resolution. In another recent study, a resolution of 5.5~mm FWHM at 662~keV has been obtained for a 30~mm thick LaBr$_3$:Ce crystal~\cite{thirolf2016}, which is a very promising result for a crystal of such thickness. In this paper, we report the position resolution measured for 81~keV and 356~keV gamma-rays using 20~mm and 10~mm thick CeBr$_3$ crystals coupled to a $4\times4$ array of silicon photomultipliers (SiPMs). We also compare the measured CeBr$_3$ resolution with our earlier results obtained with a 20~mm thick LaBr$_3$:Ce crystal~\cite{ulyanov2016}. For completeness we report the relative photoelectron yield and gamma-ray energy resolution of all three crystals measured at 662~keV. 


\section{Detector modules}
\begin{figure}[h!]
   \begin{center}
       \includegraphics[width=9cm]{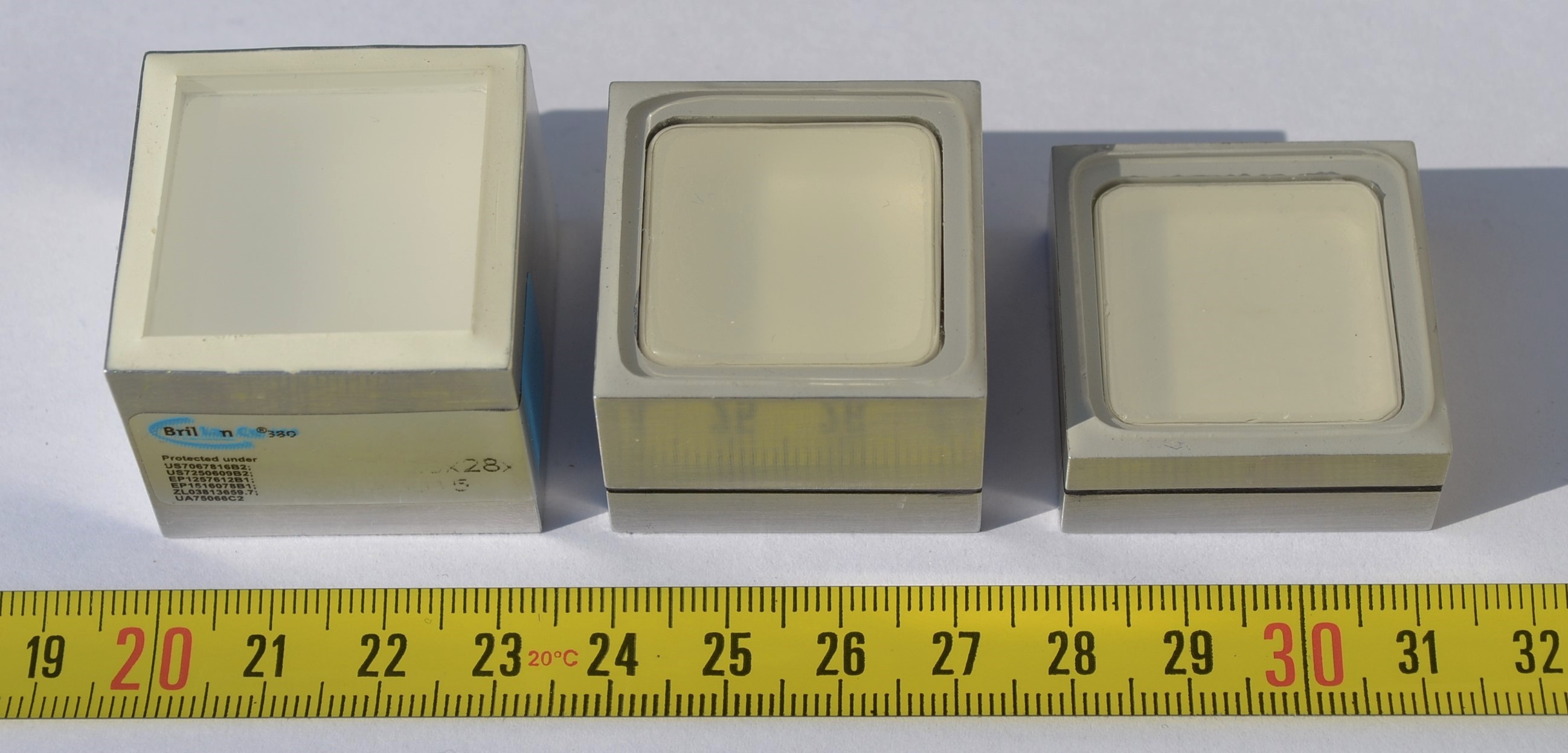}
   \end{center}
  \caption{Encapsulated scintillators: a LaBr$_3$:Ce crystal (left) and two CeBr$_3$ crystals (middle and right). The crystals are illuminated from the left side to produce a shadow on the bottom of each crystal. In the case of the LaBr$_3$:Ce crystal, the shadow is strongly blurred and barely visible due to the light scattering that occurs at the interface between the crystal and the optical window.}
\label{crystals.png}
\end{figure}

The scintillator crystals used in this study include a LaBr$_3$:Ce crystal supplied by Saint-Gobain Crystals and two CeBr$_3$ crystals supplied by SCIONIX. The supplied crystals are shown in Figure~\ref{crystals.png}. The crystal sizes are $28\times 28 \times 20$~mm$^3$ for LaBr$_3$:Ce,  $25\times 25 \times 20$~mm$^3$ and  $25\times 25 \times 10$~mm$^3$ for CeBr$_3$. All crystals are wrapped with PTFE and hermetically sealed inside aluminium containers, with an optical window on one side. The optical window of the LaBr$_3$:Ce container is 5~mm thick and matches the size of the crystal. The optical windows of the CeBr$_3$ containers are only 2~mm thick and slightly larger than the crystals, extending to  $30\times 30$~mm$^2$. The light collected from the LaBr$_3$:Ce crystal is strongly diffused at the interface between the crystal and the optical window, probably due to the rough (unpolished) surface of the crystal. Light scattering is so strong that it is hard to see the bottom and the sides of the crystal through the optical window. A highly diffusive interface like this is commonly used for large LaBr$_3$:Ce and CeBr$_3$ crystals to make the spatial distribution of the output light more uniform and thus reduce the effect of photodetector non-uniformity on light measurements. 
However, the more uniform light distribution is less dependent on the position of the gamma-ray interaction point, and the position reconstruction in this case becomes less accurate. For the supplied CeBr$_3$ crystals, the interface to the optical window has been made less diffusive and the bottom of the crystal can be seen more easily. According to SCIONIX, some diffusion of light is unavoidable as CeBr$_3$ crystals cannot be polished.

\begin{figure}[h!]
   \begin{center}
       \includegraphics[width=6cm]{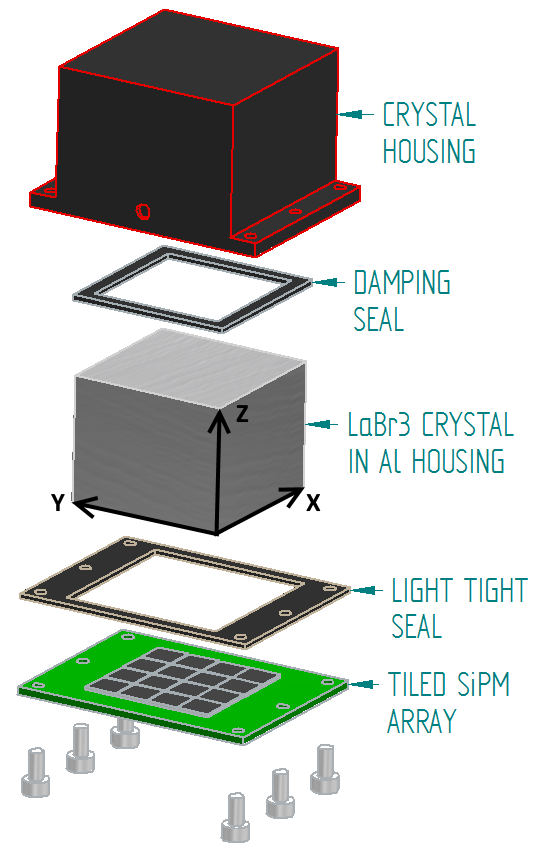}
   \end{center}
  \caption{Gamma-ray detector assembly. A 3~mm thick opaque perspex housing is used to attach the crystal to the SiPM array. }
\label{assembly.png}
\end{figure}

\begin{figure}[h!]
   \begin{center}
       \includegraphics[width=6cm]{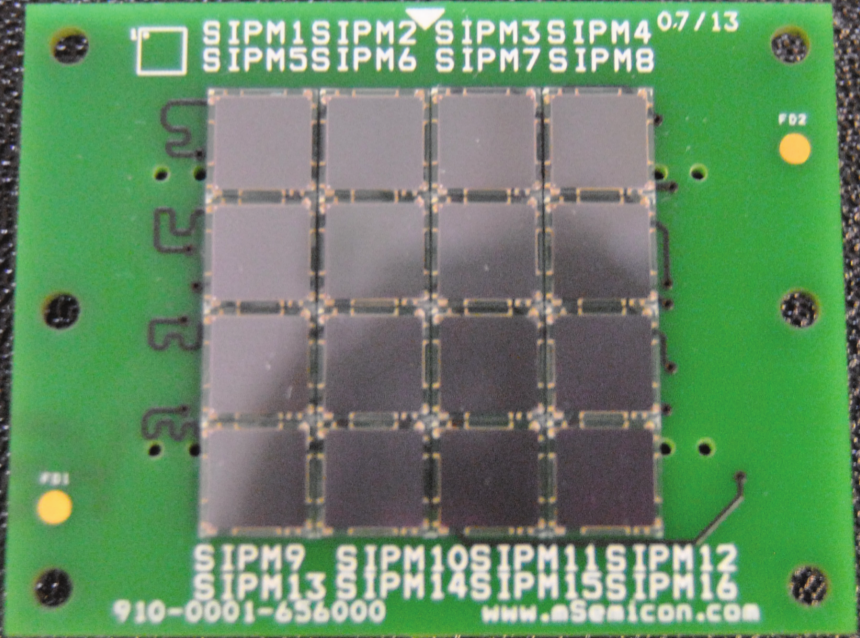}
   \end{center}
  \caption{Custom-built SiPM array.}
\label{sipmarray.png}
\end{figure}

The scintillator crystals were coupled to a $4\times4$ SiPM array as shown in Figure~\ref{assembly.png}. BC-630 optical grease from Saint-Gobain Crystals was used between the optical window of the crystal package and the SiPMs. The SiPM array is shown in Figure~\ref{sipmarray.png}. The total area of the array was $29\times29$~mm$^2$, approximately matching the crystal dimensions. The array was custom built using sixteen $6\times6$~mm$^2$ blue sensitive SiPMs supplied by SensL (MicroFB-60035-SMT). Each SiPM had 18980 microcells, yielding a total of 303680 microcells for the entire array. The SiPMs were mounted on a PCB with a pitch of 7.2~mm and were read out individually using a custom preamplifier board and two 8-channel CAEN V1720 waveform digitisers (12 bit, 250 MS/s). The on-board FPGAs performed digital signal integration. The SiPM array was operated at a bias voltage of 28~V, corresponding to 3.5~V above the breakdown voltage. The bias voltage was regulated as a function of the SiPM temperature in order to maintain constant over-voltage and thus prevent changes in the SiPM gains. 

As any fired SiPM microcells are insensitive to light and need time to recover, the SiPM response is not proportional to the number of incident photons and becomes saturated when all microcells are fired. Non-linearity of the SiPM response can be corrected using the inverse of the response function. The non-linear response function of the SiPMs used in the array was measured using short LED pulses in our previous work~\cite{ulyanov2016}. The corrected SiPM response was found to be proportional to the number of photons up to about $10^5$ photons per single SiPM, or about 2 primary avalanche triggers ("photoelectrons") per microcell (because these triggers are randomly distributed among the microcells, about 13\% of the microcells still remain unfired for such light pulses). Although the pulse height resolution of the SiPMs deteriorated at high light levels, the resolution of each single SiPM was measured to be better than 3\% FWHM for light pulses of $10^5$ photons. The 16 pixel array should be able to accurately measure scintillation light pulses produced in LaBr$_3$:Ce by 50~MeV gamma-rays. 
The SiPM array coupled to a LaBr$_3$:Ce crystal was tested with gamma-rays over an energy range of 30~keV to 9.3~MeV. The detector response corrected for the SiPM non-linearity was found to be proportional to the gamma-ray energy, except for the known LaBr$_3$:Ce non-proportionality below 100~keV~\cite{khodyuk2012, quarati_cebr3}. For 356~keV gamma rays used in this study, only about 1\% of the SiPM microcells are fired, therefore the SiPM
 non-linearity can be ignored.


\section{Position resolution measurements}

Localisation of the gamma-ray interaction point in the scintillator crystal was performed using artificial neural networks (ANNs). For each detector, three separate ANNs were implemented to calculate the $x$, $y$ and $z$-coordinates of the interaction point based on the signals of the 16 SiPM pixels. The coordinate axes are defined as shown in Figure~\ref{assembly.png}. The ratios of individual SiPM outputs to the total signal of the detector were used as ANN inputs. As the sum of all 16 signal ratios always equals unity, only 15 independent signal ratios (pixel 1 to 15) were used for input. The ANNs were implemented using the Toolkit for Multivariate Data Analysis~\cite{tmva} (TMVA version 4.2.0) included in the ROOT data analysis framework version 5.34~\cite{root}. 

\begin{figure}[h!]
   \begin{center}
       \includegraphics[width=9cm]{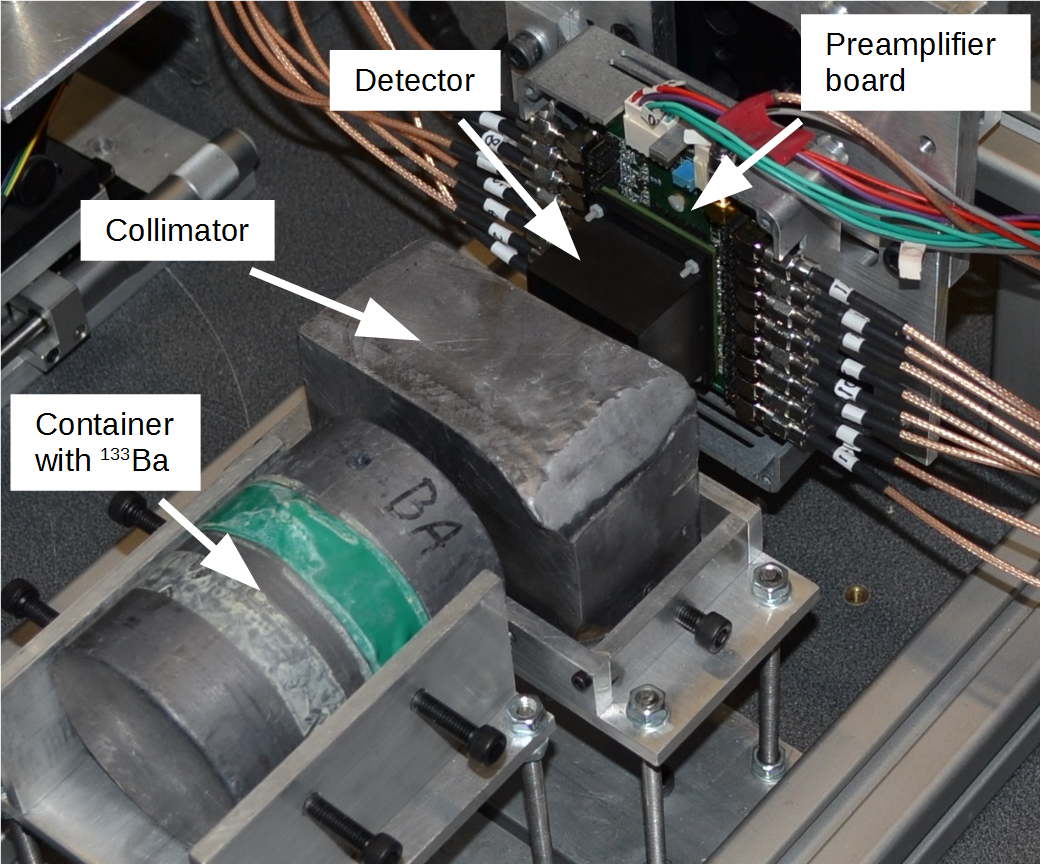}
   \end{center}
  \caption{Experimental set-up for position resolution measurements. The detector is mounted vertically and the gamma-ray beam hits the front face of the crystal. The detector can also be positioned horizontally, which allows a side face of the crystal to be irradiated.}
\label{beamline}
\end{figure}

\begin{figure}[h!]
   \begin{center}
       \includegraphics[width=9cm]{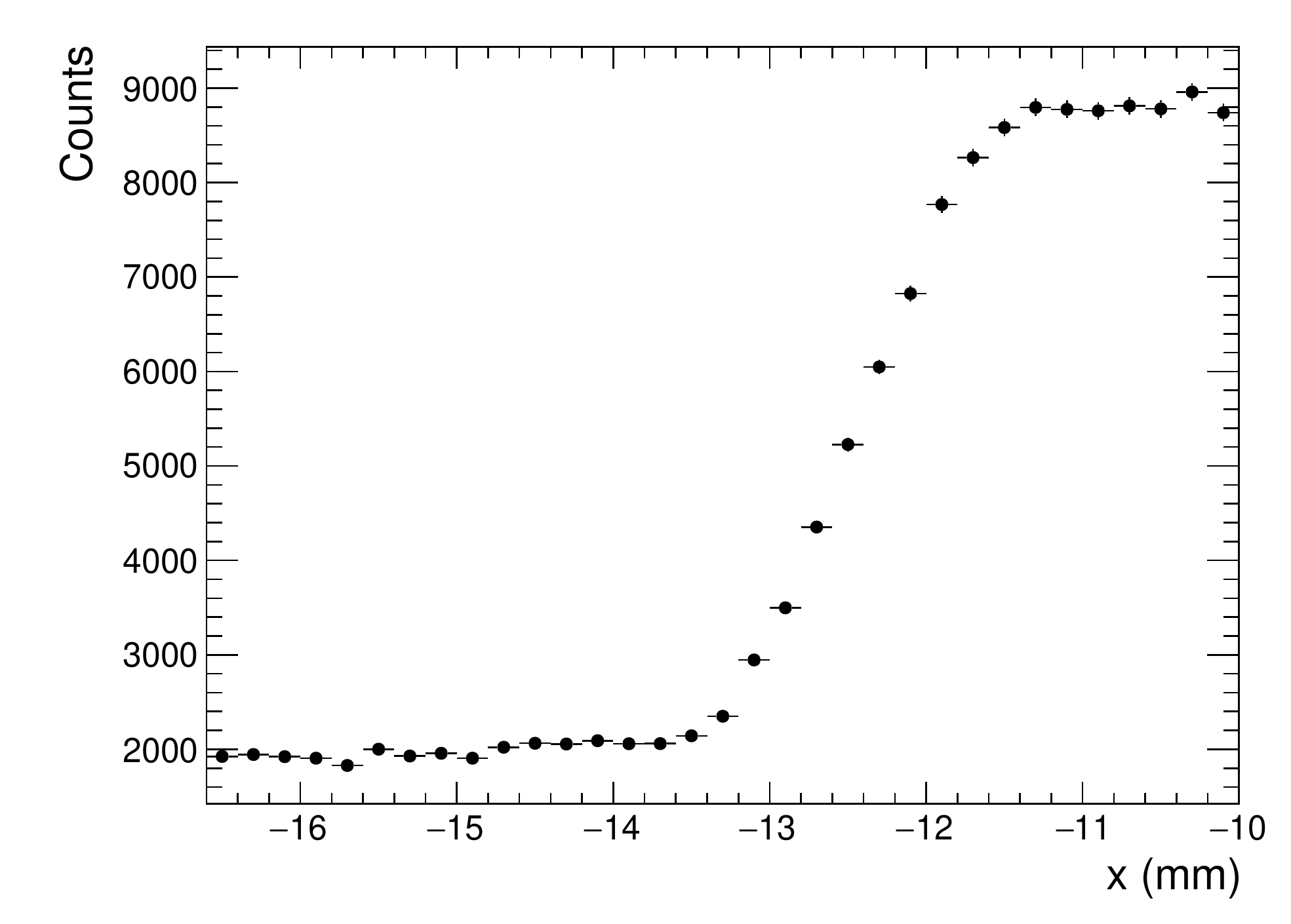}
       \includegraphics[width=9cm]{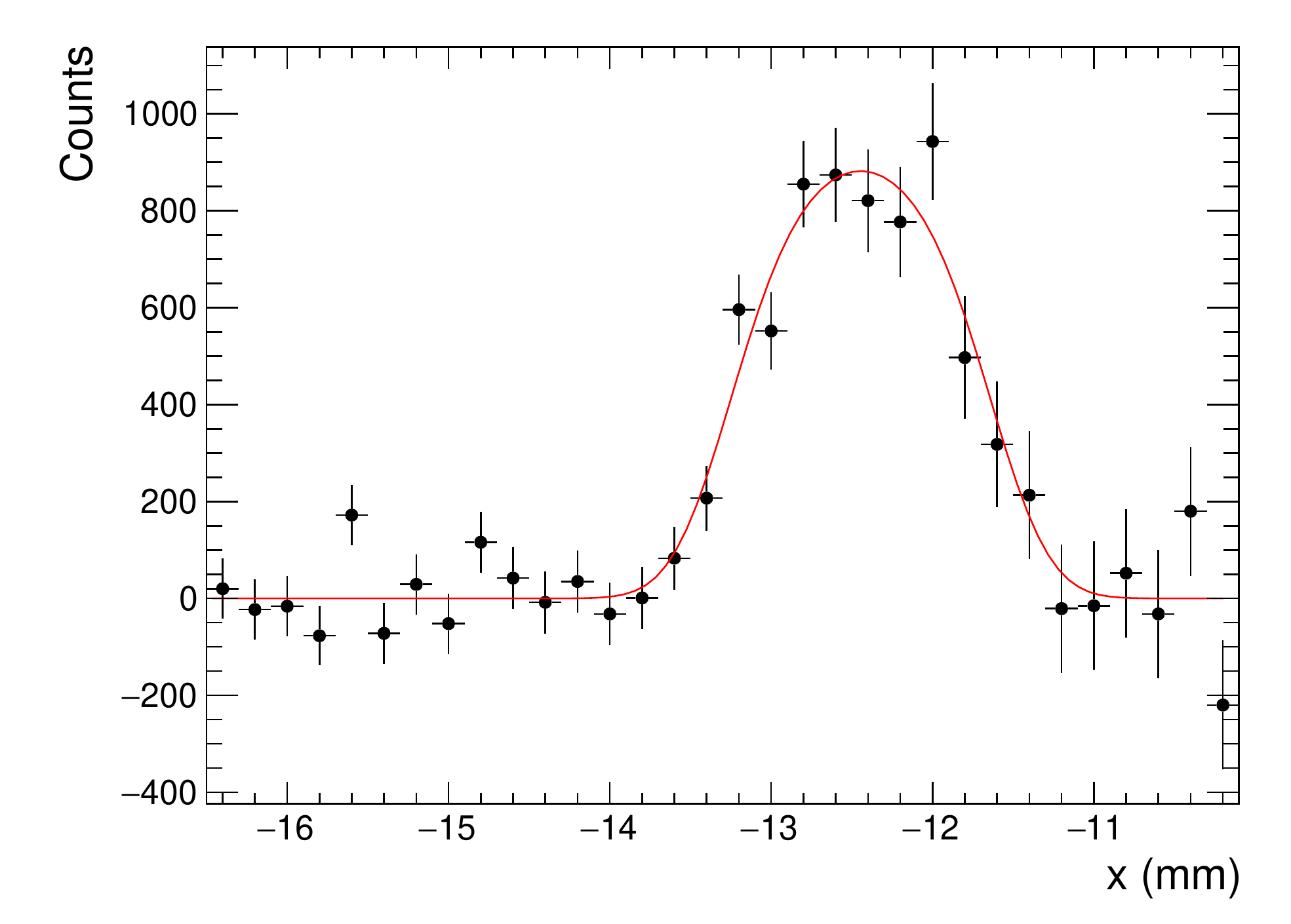}
   \end{center}
  \caption{Top: Event rate as a function of the beam position with respect to the detector. The rate increases from the background level to the full rate as the scintillator crystal moves into the beam. The data acquisition time for each position is 1000~s. 
  Bottom: Beam profile. The data points are increments of the event rate per position step $\Delta x = 0.2$~mm. The curve is the profile of a uniform circular beam convolved with a Gaussian function that was fitted to the data. 
  }
\label{beamprof}
\end{figure}

To collect data for ANN training, the assembled detector module was mounted on horizontal and vertical linear stages and scanned with a collimated $^{133}$Ba source, as shown in Figure~\ref{beamline}. The collimator was a 5 cm thick lead brick with a 1.4~mm aperture. The width of the gamma-ray beam was estimated experimentally by positioning the 10~mm thick CeBr$_3$ detector near the beam line and measuring the event rate as the detector was moved into the beam. The beam profile measured along the X-axis is shown in Figure~\ref{beamprof}. Similar results were obtained when the beam crossed the other crystal edge and for the measurements along the Y-axis. The full width of the beam was found to be about 1.6~mm at half maximum.

\begin{figure}[h!]
   \begin{center}
       \includegraphics[width=9cm]{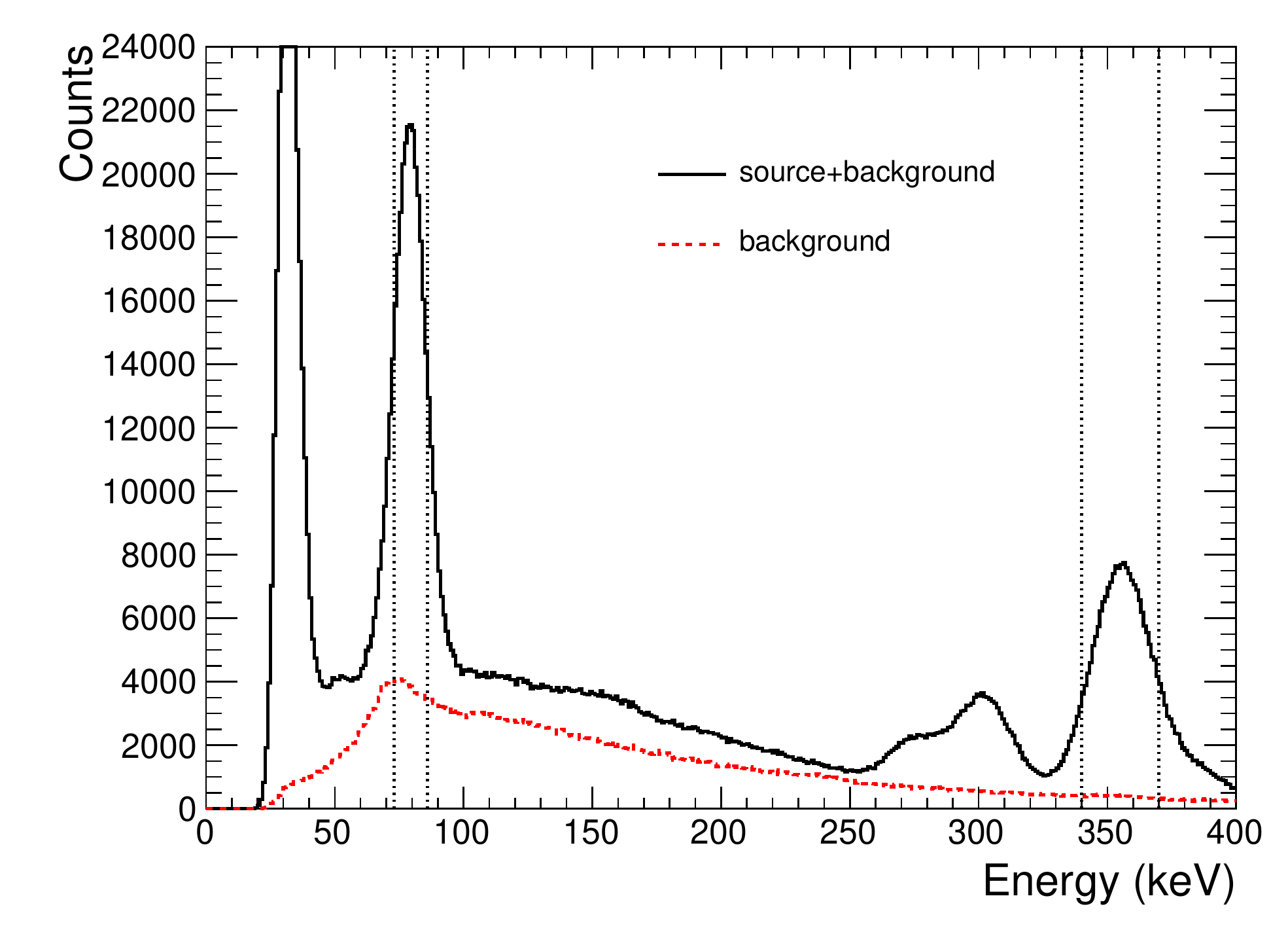}
   \end{center}
  \caption{Spectrum of the collimated $^{133}$Ba source measured with the 20~mm thick CeBr$_3$ detector. The spectrum of background radiation (measured without the source using equal acquisition time) is shown for comparison. The vertical dotted lines show the energy bands used to select gamma-rays consistent with the 81~keV and 356~keV lines.}
\label{ba133spectrum20xy.pdf}
\end{figure}

The detector module was first mounted vertically and the front face of the detector was scanned across the beam (XY scan). Then, the module was rotated and a side of the detector was scanned (XZ scan). The XY scan was performed using a grid of $11 \times 11$ positions with a step of 2.3~mm to cover the entire face of the CeBr$_3$ crystal. The XZ scan was performed using a grid of $11\times 9$ positions. The step along the $z$-axis was 2.3~mm for the 20~mm thick crystal and 1~mm for the 10~mm thick crystal. For each triggered event, the beam position and the 16 SiPM signals were recorded. The data collected from the XY scan were used to train the ANNs for reconstruction of $x$ and $y$-coordinates, as the $x$ and $y$ of the gamma-ray interaction points were approximately given by the beam positions. Similarly, the XZ scan was used to train the ANN for $z$ reconstruction.

\begin{figure}[h!]
   \begin{center}
       \includegraphics[width=9cm]{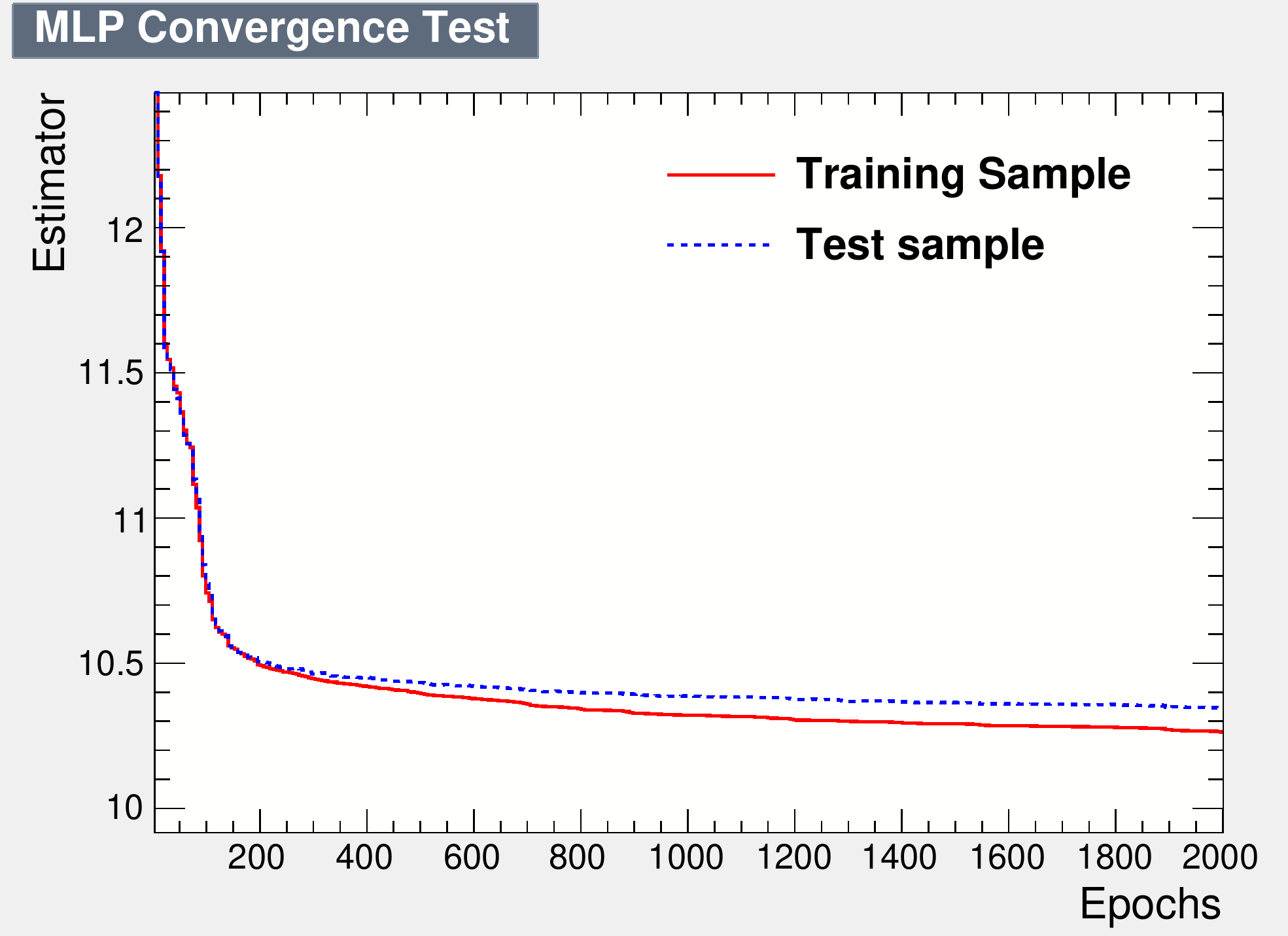}
   \end{center}
  \caption{Mean squared error (mm$^2$) for the $x$ reconstruction in the training and testing sample as a function of the number of epochs in the training process.}
\label{ANNconverge}
\end{figure}

The ANN training was performed using 356~keV gamma-rays selected as shown in Figure~\ref{ba133spectrum20xy.pdf}. The events selected from each scan were split into a training sample and a testing sample, each containing about 50,000 events. The ANNs used in this study were multilayer perceptrons with the sigmoid activation function $\alpha(x)= 1/(1+e^{-x})$. Each ANN had a single hidden layer of 10 neurons, which was found to be sufficient for optimal position reconstruction. The Broyden-Fletcher-Goldfarb-Shannon (BFGS) method was employed for ANN training and the mean squared error was used as the ANN cost function. The ANNs were trained  using 2000 epochs, although training beyond the first 200 epochs gave very little improvement in the ANN performance as can be seen in Figure~\ref{ANNconverge}.   

In order to validate the reconstruction performance of the trained ANNs, additional XY and XZ scans were carried out using different irradiation positions. The XY scan was performed using a grid of $24 \times 24$ positions with a step of 1~mm. The XZ scan was performed using a $24\times 20$ grid for the 20~mm thick crystal and a  $24\times 10$ grid for the 10~mm thick crystal. From theses scans, the $x$, $y$ and $z$-position resolution was assessed for 81~keV and 356~keV gamma-rays.

For each gamma-ray in the performed position scans, two coordinates of the interaction point were known from the beam position. Therefore, a single ANN with two outputs (for example, $x$ and $y$) could be trained as an alternative to two ANNs with single outputs. Provided a sufficient number of neurons are used in the hidden layers, these two approaches should be equivalent in terms of the achievable reconstruction performance. Let us consider an arbitrary ANN with two output neurons and let $x_\mathrm{rec}(\mathrm{input})$ and $y_\mathrm{rec}(\mathrm{input})$ be the functions (defined by the ANN weights) that describe the dependence of the two outputs on the input variables. In this case the mean-squared-error cost function is given by 
\begin{equation}
\overline{\Delta r^2} = \overline{\Delta x^2 + \Delta y^2} = \overline{\Delta x^2} + \overline{\Delta y^2},   
\label{eq_mse}
\end{equation}
where $\Delta x = x_\mathrm{rec}-x_\mathrm{true}$  and $\Delta y = y_\mathrm{rec}-y_\mathrm{true}$ are deviations of the ANN outputs (estimators) from the true parameter values and the bars designate the sample mean values. Let $\overline{\Delta x_0^2}$ and $\overline{\Delta y_0^2}$ be the mean squared deviations corresponding to the optimal ANN weights that minimise the cost function (\ref{eq_mse}). Now consider two ANNs  with the same topology as the network above but with a single output neuron. Using the weights identical to the ANN with two outputs, these two networks can provide the same response functions $x_\mathrm{rec}(\mathrm{input})$ and $y_\mathrm{rec}(\mathrm{input})$ and the same mean squared deviations. The weights of the two networks are independent, therefore they can represent a larger set of response functions than the ANN with two outputs. Since each ANN is trained to minimise the mean squared deviation for one parameter, these two networks should perform better than the ANN with two outputs, yielding the mean squared deviations $\overline{\Delta x_1^2} \le  \overline{\Delta x_0^2} $  and   $\overline{\Delta y_1^2} \le  \overline{\Delta y_0^2}$. As follows from Equation (\ref{eq_mse}), the mean squared error for two parameters is also reduced:  $\overline{\Delta r_1^2} \le  \overline{\Delta r_0^2}$. It should be noted, however, that two ANNs with single outputs can always be combined into one larger ANN with two outputs. In that case, the ANN with two outputs can represent a larger set of response functions and should have a smaller mean squared error after training than the two smaller networks. The performance of ANNs of one type can always be matched or improved by ANNs of the other type. If the number of neurons in the hidden layers is sufficiently large, the difference between the two types of ANNs becomes insignificant.

\section{Relative light yield and energy resolution measurements}

The relative photoelectron yield and energy resolution were measured using an uncollimated mixed source containing $^{137}$Cs and $^{60}$Co radioisotopes. The source was placed in front of the detector at a distance of several millimetres and a spectrum of the detector response was recorded. The photoelectron yield was calculated relative to the LaBr$_3$:Ce crystal using the positions of the 662~keV photoabsorption peaks in the acquired spectra. The widths of the 662~keV peaks were used to calculate the energy resolution of the detectors. The measurements were performed using the SiPM array and then repeated using a Hamamatsu R6233-100 photomultiplier tube (PMT), which is a 3-inch diameter 8-stage tube with a super bialkali photocathode. The spectral sensitivity of this PMT is well suited for detection of LaBr$_3$:Ce and CeBr$_3$ scintillation emission, with the quantum efficiency reaching a maximum value of 35\% at 350~nm.
The PMT was supplied with a tapered voltage divider (Hamamatsu) to improve signal linearity for large peak currents and was operated at a bias voltage of 1000~V. 
The output of the PMT was connected directly to the V1720 digitiser, which sampled and digitally integrated the signals. The signal integration time was set to 140~ns in the PMT measurements. The integration time for the SiPM signals was set to 400~ns to accommodate the relatively long pulses from the SiPM array. The long decay time of the signals is due to the large capacitance of the SiPMs used in the array. Smaller SiPMs found in many applications typically have much shorter pulses.  

\section{Results}
\subsection{Position resolution}
\begin{figure}[!h]
  \begin{center}
    \includegraphics[width=6cm]{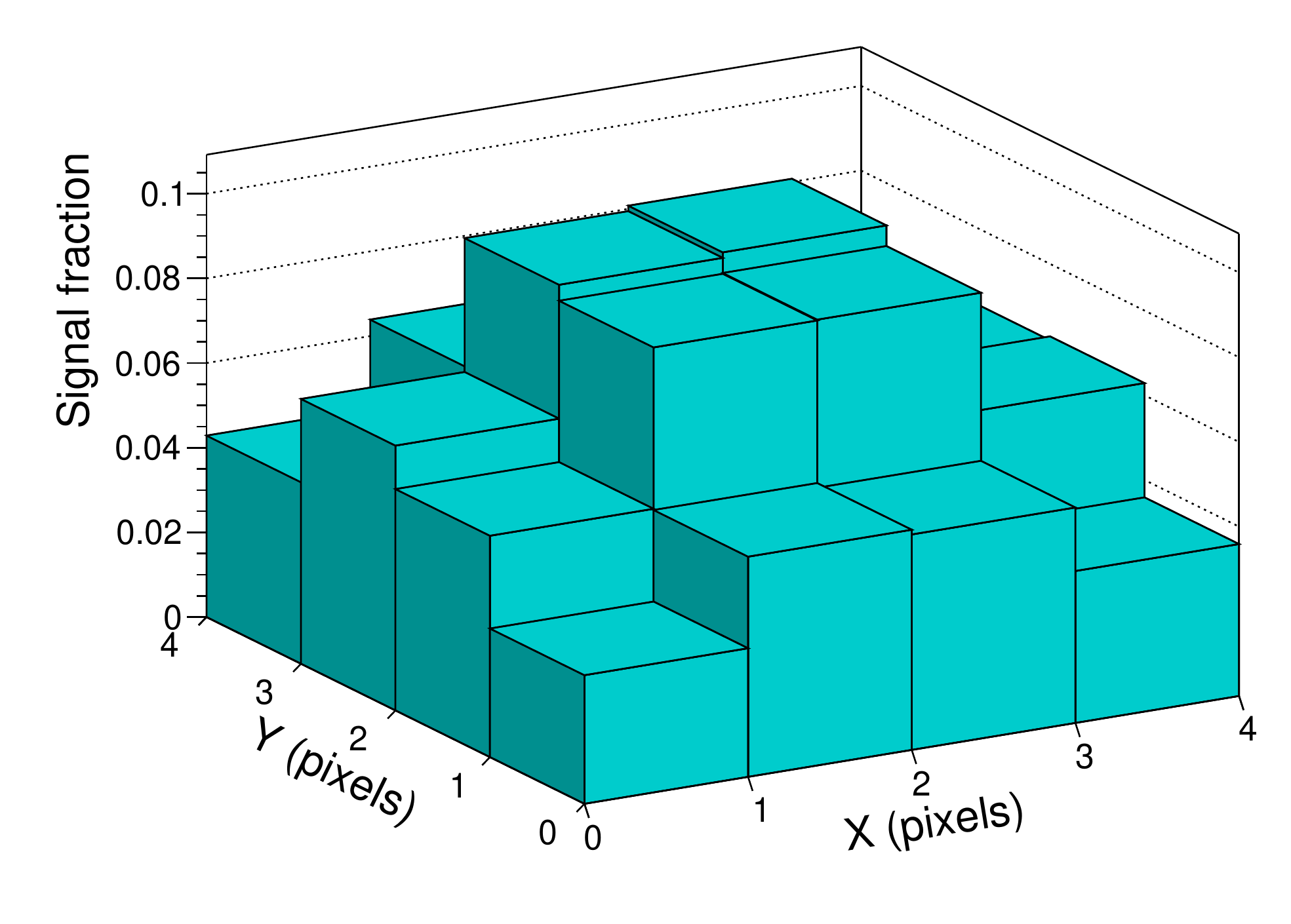}
    \includegraphics[width=6cm]{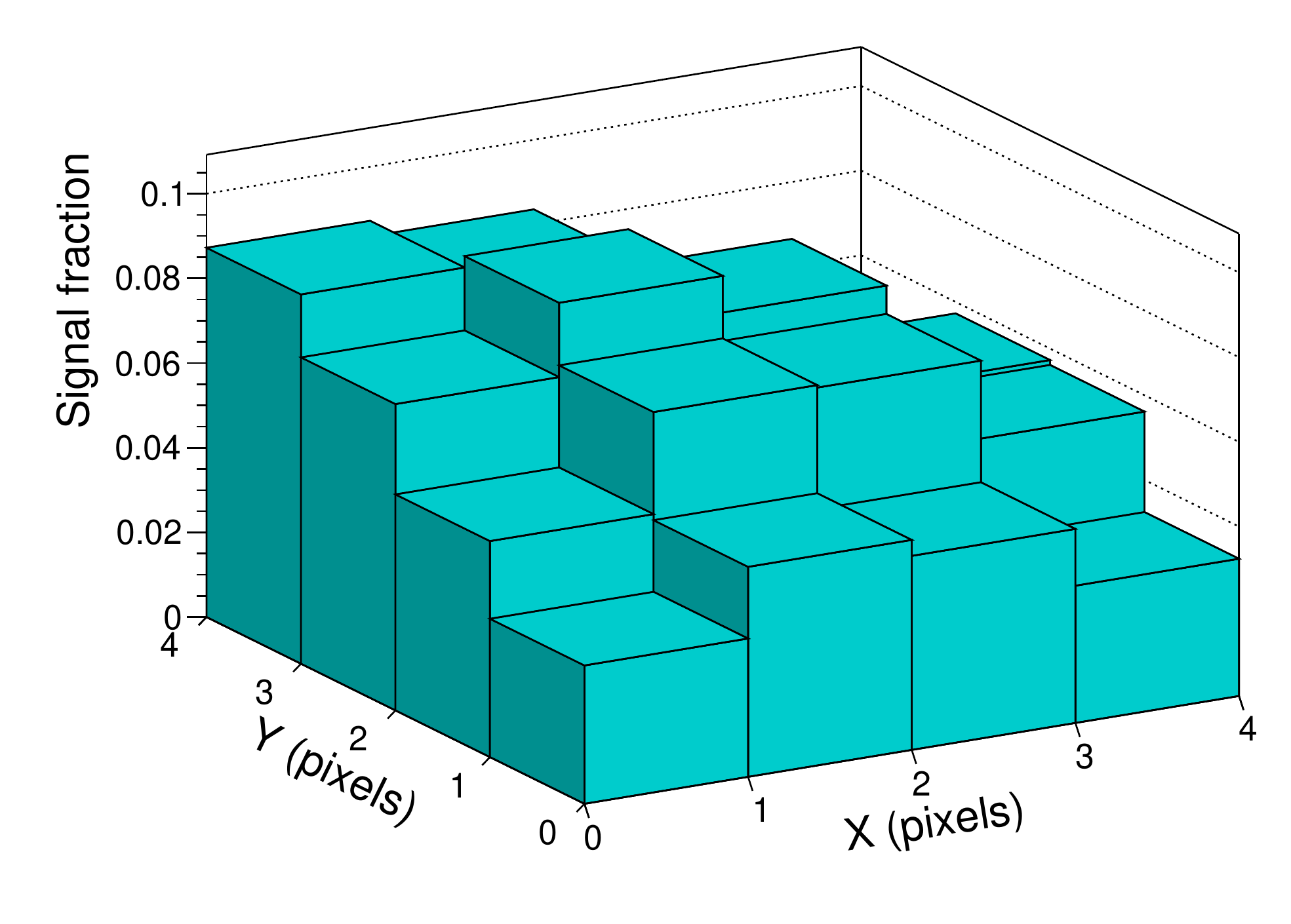}
    \includegraphics[width=6cm]{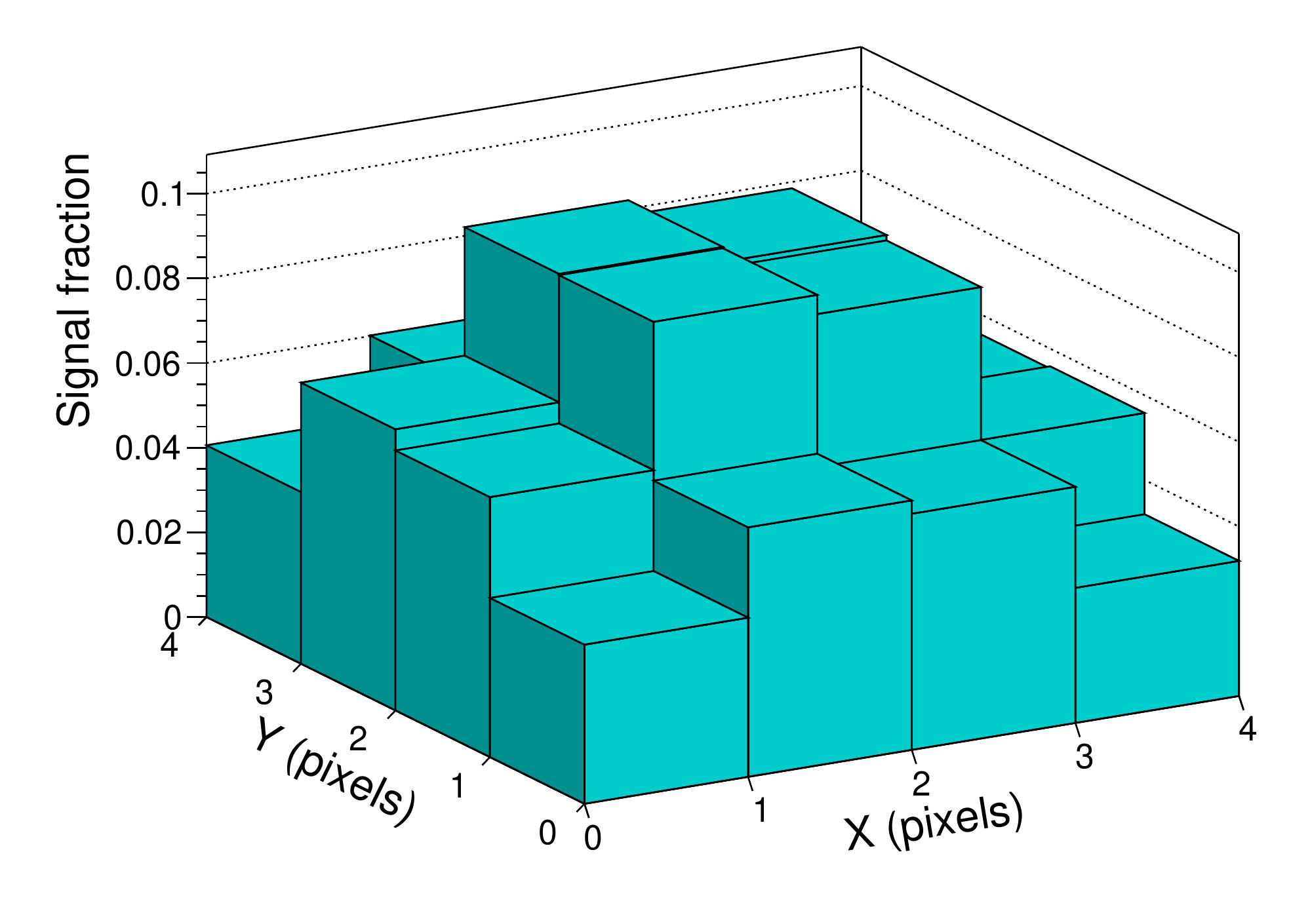}
    \includegraphics[width=6cm]{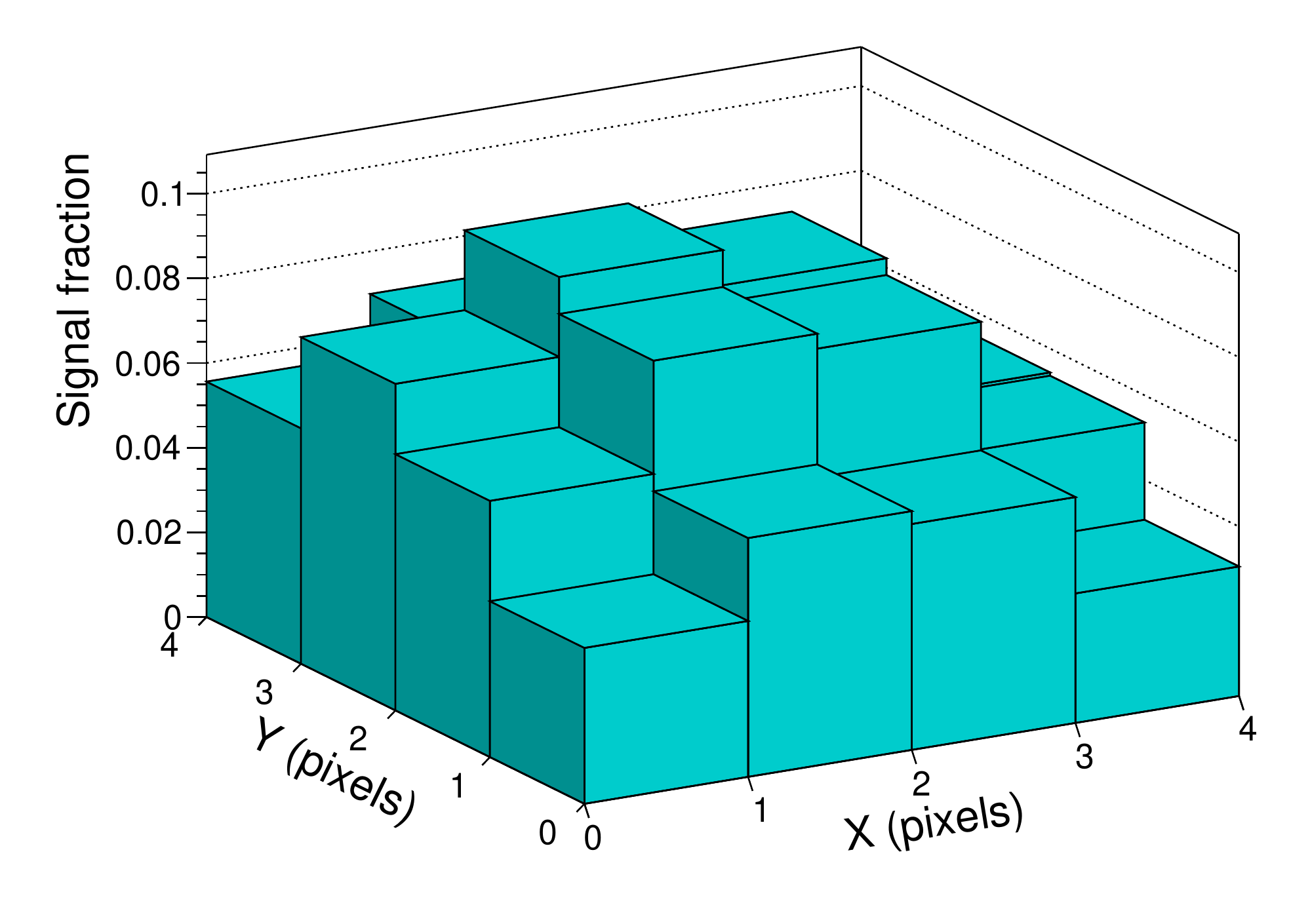}
    \caption{Average signal distributions for fully absorbed 356~keV gamma-rays from the collimated $^{133}$Ba source.  The gamma-ray beam points near the centre of the crystal ($x_\mathrm{beam}=-0.5$~mm, $y_\mathrm{beam}=0.5$~mm, left plots) and at the corner of the crystal ($x_\mathrm{beam}=-11.5$~mm, $y_\mathrm{beam}=11.5$~mm, right plots). The upper plots are for the 10~mm thick CeBr$_3$ crystal and the bottom plots are for the 20~mm thick CeBr$_3$ crystal. }
    \label{pxl_dist}
  \end{center}
\end{figure}

 The average signal distribution among the SiPM pixels is shown in Figure~\ref{pxl_dist} for two positions of the gamma-ray beam. The signal distribution clearly depend on the position of the beam and can therefore be used for reconstruction of the gamma-ray interaction points. The effect of the beam position on the signal distribution for the thicker crystal is much smaller, which means the position reconstruction cannot be done as accurately as in the thin crystal. The signal distributions for individual events are subject to statistical fluctuations, mostly defined by the number of photoelectrons generated in the SiPMs. Together with the signal dependence on the interaction position, these fluctuations define the position resolution of the detector.

\begin{figure}[!h]
  \begin{center}
    \includegraphics[width=9cm]{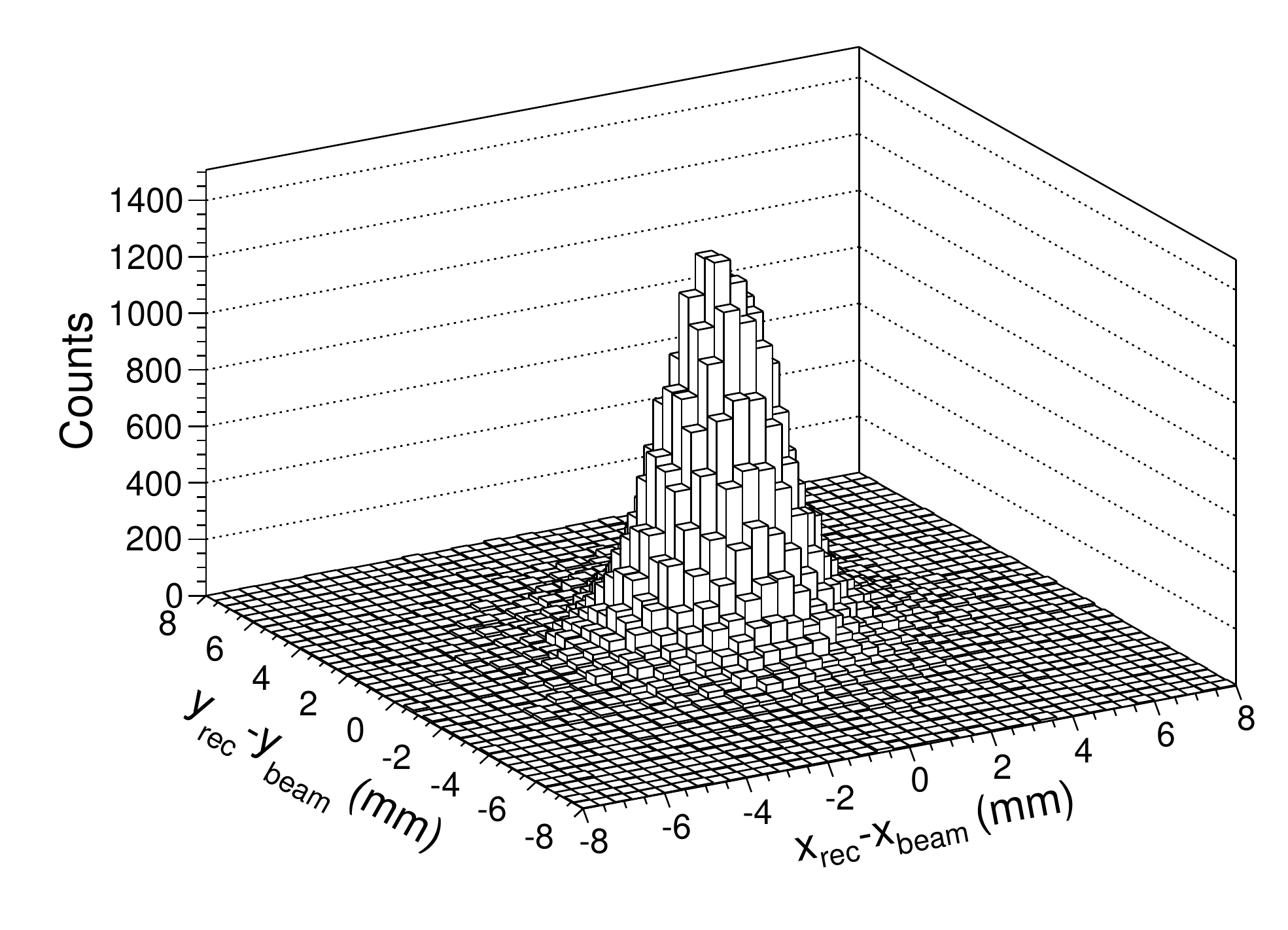}
    \includegraphics[width=9cm]{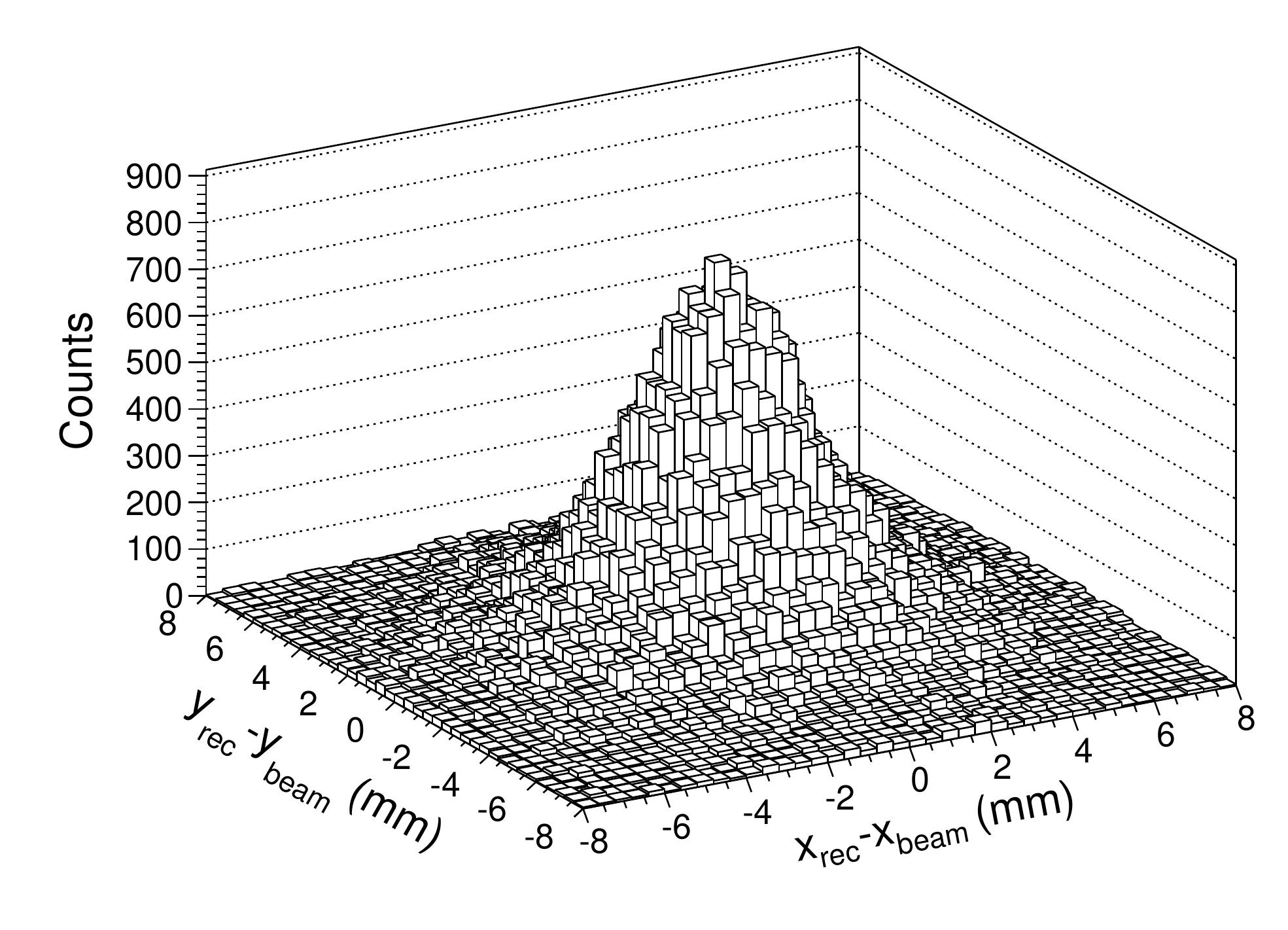}
    \caption{Average point spread function for 356~keV gamma-rays for the 10~mm thick CeBr$_3$ detector (top) and for the 20~mm thick CeBr$_3$ detector (bottom). 
    }
    \label{deltaxy.pdf}
  \end{center}
\end{figure}

\begin{figure}[!h]
  \begin{center}
    \includegraphics[width=9cm]{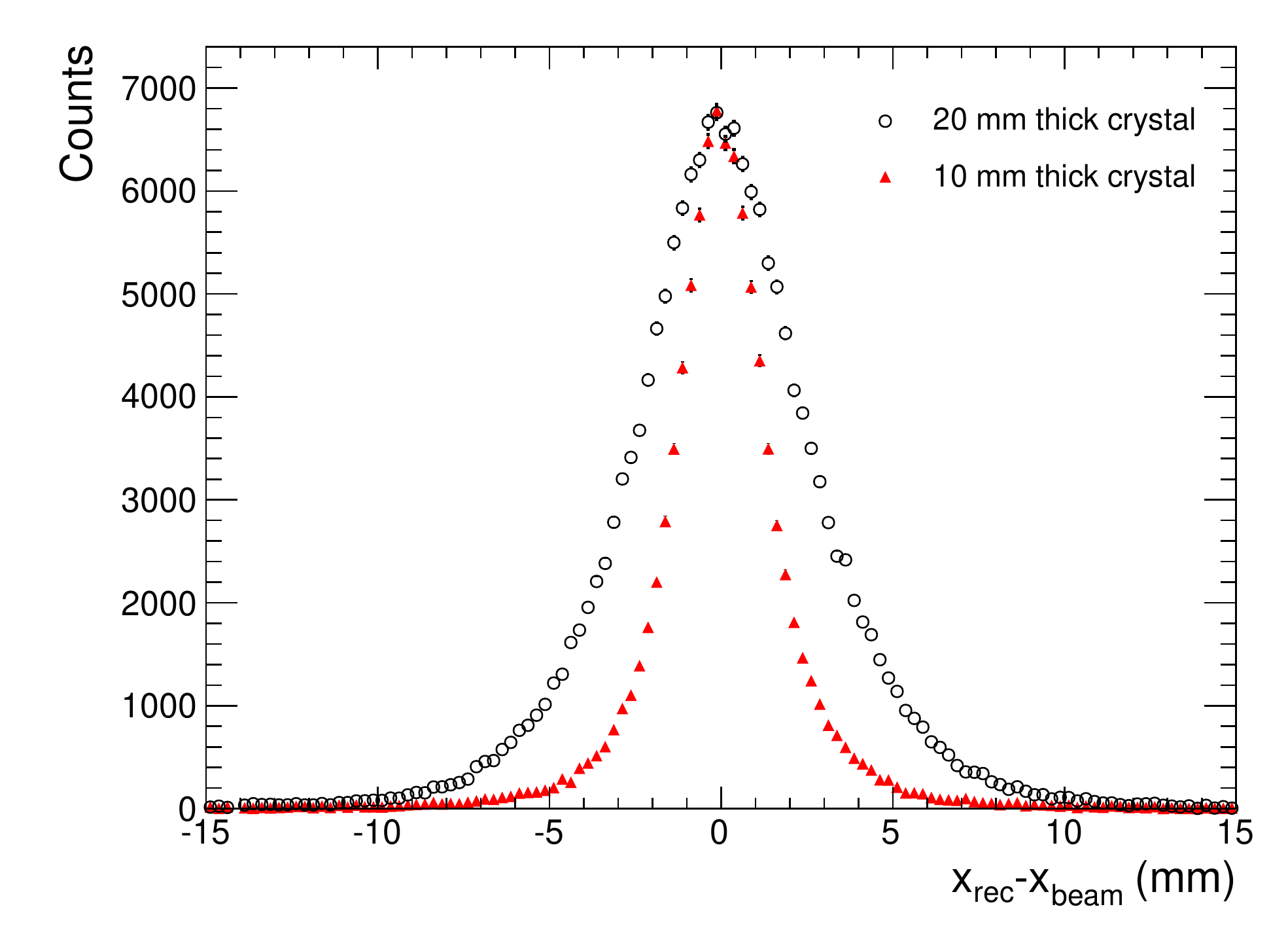}
    \includegraphics[width=9cm]{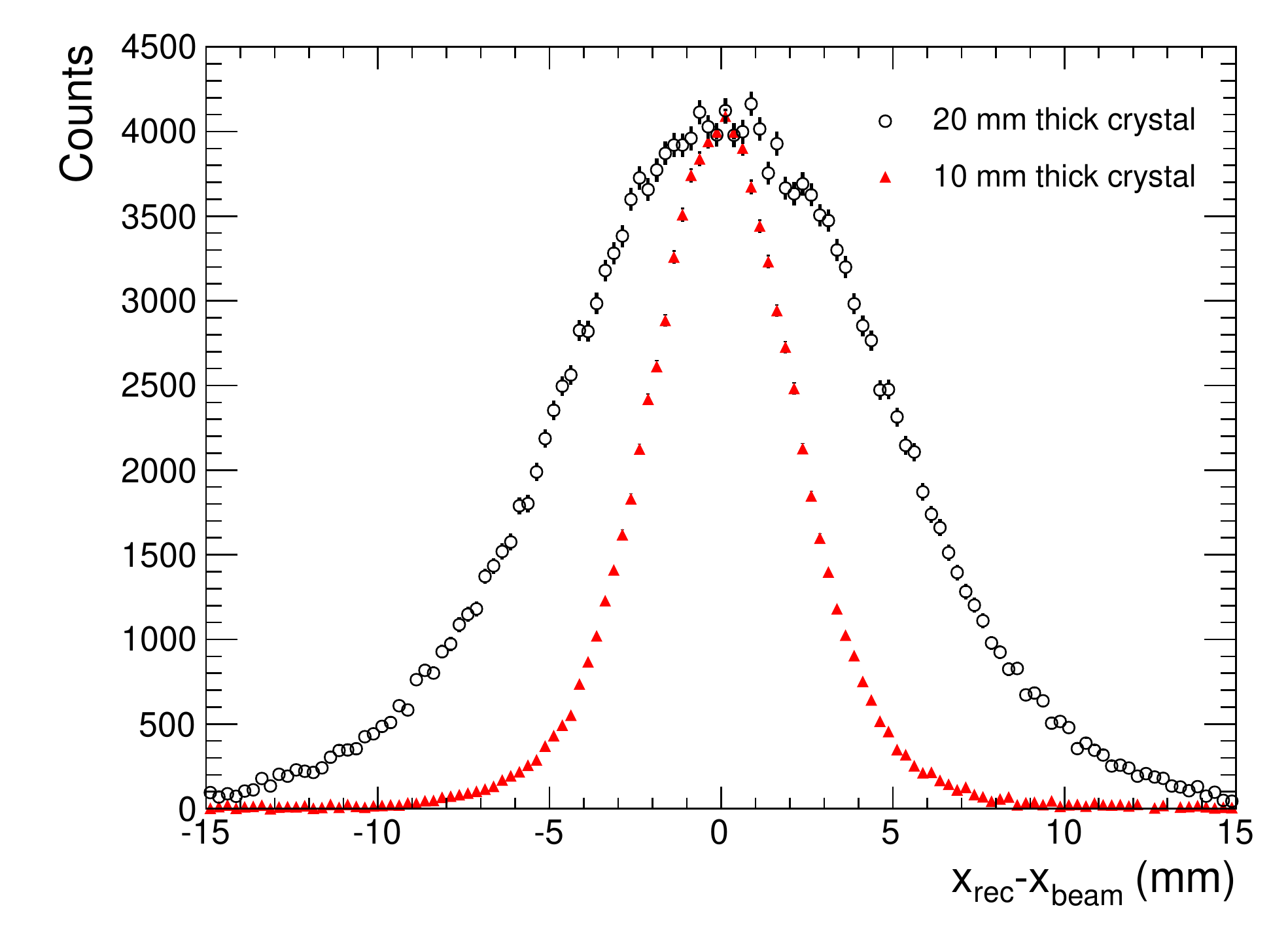}
    \caption{Statistical distributions of the position reconstruction error $x_\mathrm{rec}-x_\mathrm{beam}$, obtained from the XY position scans of the two CeBr$_3$ detectors with 356~keV gamma-rays (top) and 81~keV gamma-rays (bottom). The distributions for the 10~mm detector were scaled to equalise the heights of the distributions for the two detectors. The background distributions (measured using identical scans and the same acquisition time but without source) have been subtracted.}
    \label{deltax.pdf}
  \end{center}
\end{figure}

\begin{figure}[!h]
  \begin{center}
    \includegraphics[width=9cm]{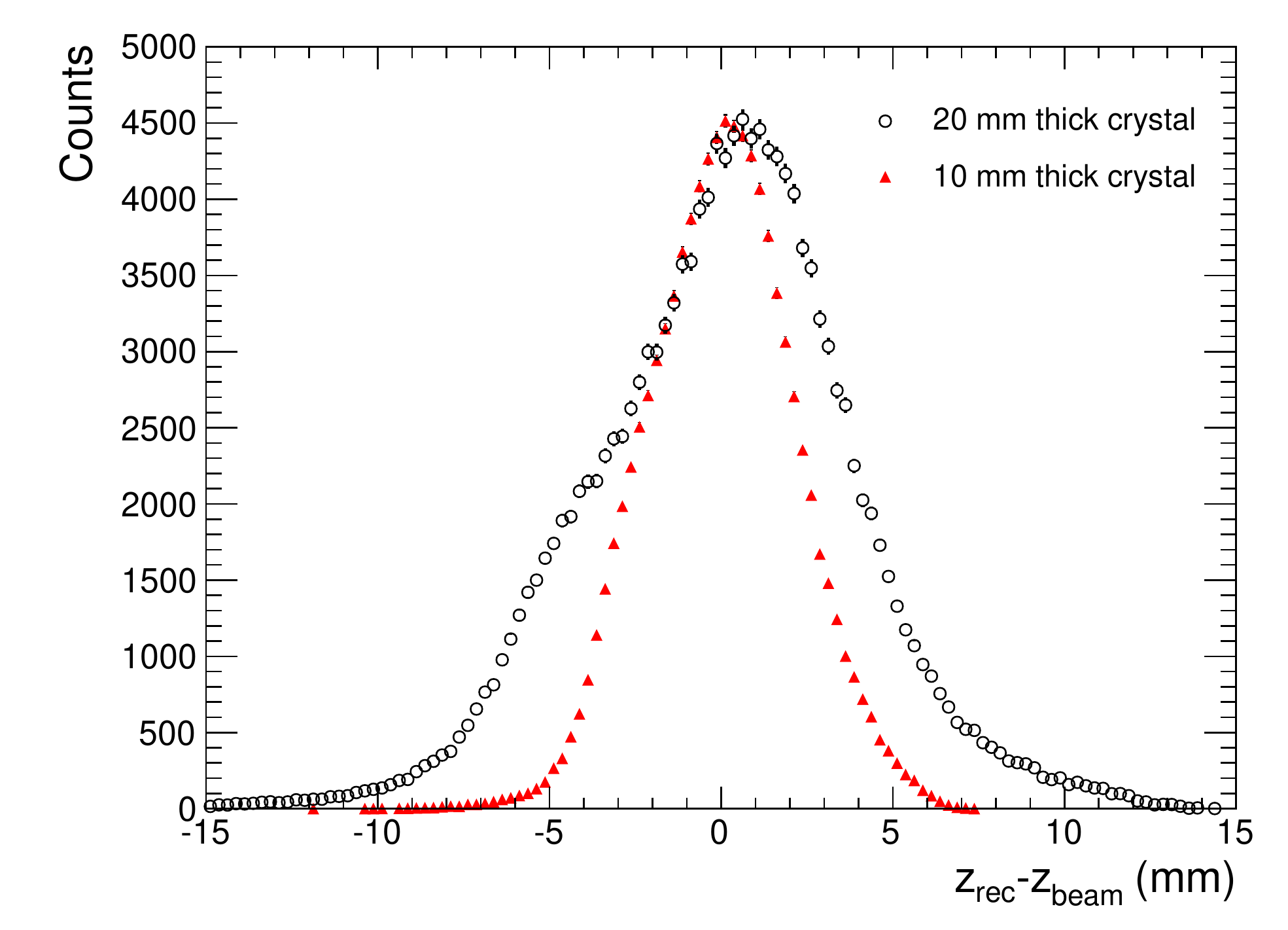}
    \includegraphics[width=9cm]{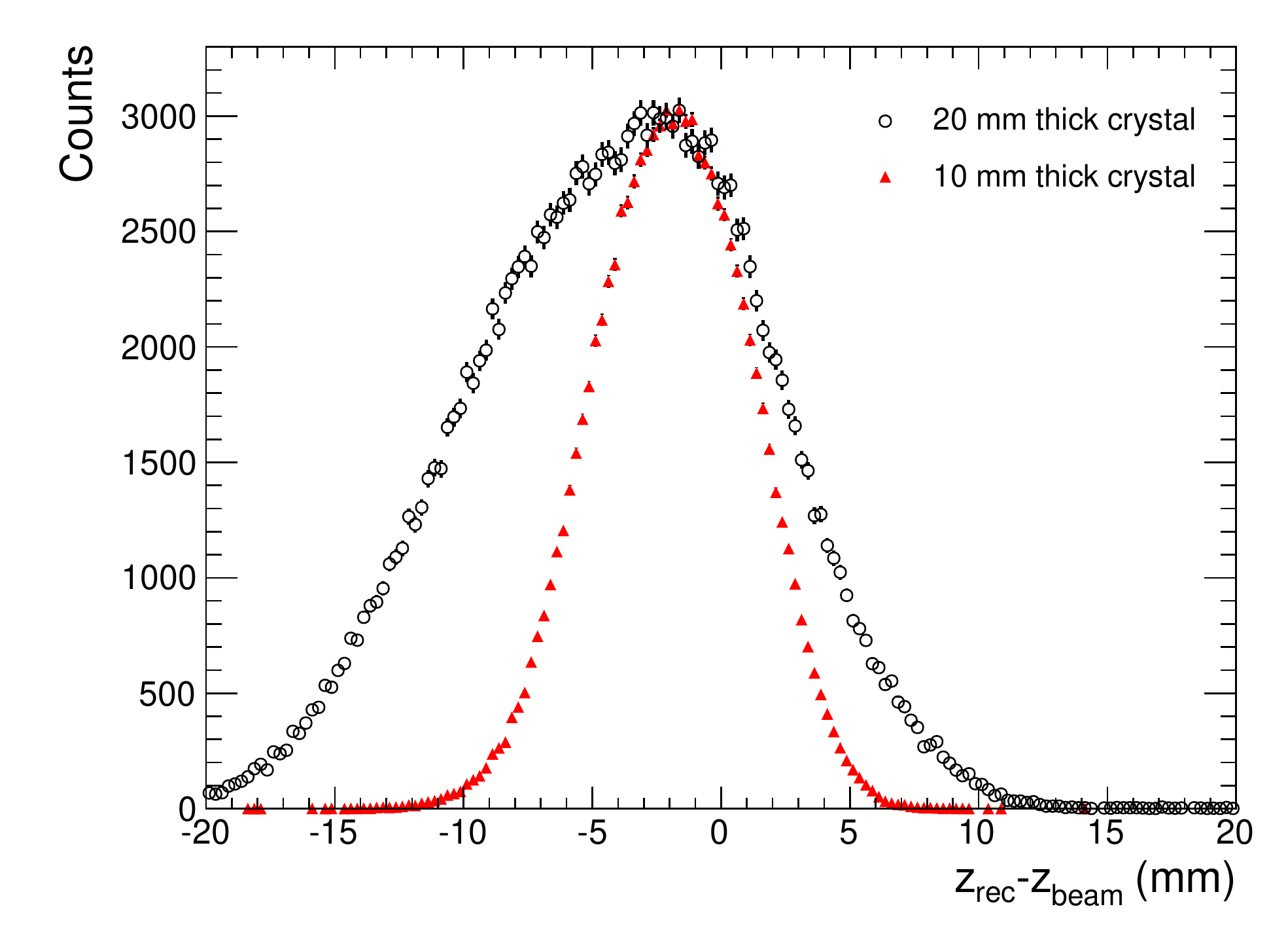}
    \caption{Statistical distributions of the position reconstruction error $z_\mathrm{rec}-z_\mathrm{beam}$, obtained from the XZ position scans of the two CeBr$_3$ detectors with 356~keV gamma-rays (top) and 81~keV gamma-rays (bottom). The background distributions have been subtracted.}
    \label{deltaz.pdf}
  \end{center}
\end{figure}

The average point spread function (PSF) obtained from a $24\times 24$ position scan of the entire front face of the detector is shown in Figure~\ref{deltaxy.pdf} for the two CeBr$_3$ detectors. As expected, the position reconstruction is less accurate for the thicker scintillator, resulting in a much wider PSF. Here and below, $x_\mathrm{rec}$ and $y_\mathrm{rec}$ refer to  the interaction point coordinates reconstructed by the ANNs from the SiPM signals in any given event, and $x_\mathrm{beam}$ and $y_\mathrm{beam}$ represent the beam position. Although it has a small contribution from the beam spread, the difference between the reconstructed position and the beam position (e.g. $x_\mathrm{rec}-x_\mathrm{beam}$) is further referred to as the position reconstruction error (e.g. $x$-reconstruction error). As the position reconstruction error fluctuates from event to event, the performance of position reconstruction is characterised by the statistical distribution of the position reconstruction error.  The distributions of the $x$-reconstruction error are shown in Figure~\ref{deltax.pdf} for the two CeBr$_3$ detectors.  The reconstruction error distributions obtained for the $y$-coordinate are very similar. The distributions of $z$-reconstruction error were obtained by scanning the XZ sides of the detectors and are shown in Figure~\ref{deltaz.pdf}. The FWHMs of these distributions are used as a measure of the position resolution and are summarised in Table~\ref{table_presolution}, together with our earlier results obtained with the LaBr$_3$:Ce crystal~\cite{ulyanov2016}.

\begin{table}[!h]
\caption{Average position resolution in the $x$, $y$ and $z$-directions measured for the three detectors with 81~keV and 356~keV gamma-rays. The resolution values include the contribution from the finite width of the gamma-ray beam (FWHM=1.6~mm).}
\centering
 \makebox[\linewidth]{
 \begin{tabular}{|l|ccc|ccc|} \hline
Crystal size (mm) & \multicolumn{3}{|c|}{FWHM at 356~keV} & \multicolumn{3}{|c|}{FWHM at 81~keV} \\
  & $\Delta x$ (mm)   & $\Delta y$ (mm)   & $\Delta z$ (mm) 
   & $\Delta x$ (mm)   & $\Delta y$ (mm)   & $\Delta z$ (mm)  \\
\hline
LaBr$_3$:Ce 28x28x20 & 8.0 & 8.0 & 11.0 & \multicolumn{3}{c|}{not measured} \\
CeBr$_3$ 25x25x20 & 5.4 & 5.3 & 7.8 & 10.9 & 10.8 & 13.9\\
CeBr$_3$ 25x25x10 & 2.9 & 2.9 & 5.2 & 5.0 & 5.0 &  7.6\\

\hline

\end{tabular}
}
\label{table_presolution}
\end{table}  

For the same scintillator thickness of 20~mm, the position resolution of the CeBr$_3$ detector is significantly better than that of the LaBr$_3$:Ce detector. This difference is even more striking if we take into account that the LaBr$_3$:Ce crystal generates about 40\% more light than the CeBr$_3$ crystal. The better position resolution of the CeBr$_3$ detector is possibly explained by the thinner 
optical window of the scintillator package (2~mm versus 5~mm) and the less diffusive interface between the optical window and the scintillator crystal. 

Both monolithic CeBr$_3$ detectors display sub-pixel position resolution at 356 keV, and therefore perform better than a detector using segmented scintillator with one-to-one coupling of scintillator pixels to the SiPMs. Although the position resolution of the monolithic detectors gets considerably worse for low-energy gamma-rays, the resolution of the tested CeBr$_3$ detectors at 81~keV is still acceptable for use in a Compton telescope. The degradation of the position resolution at low energies is explained by larger statistical fluctuations in the distribution of the scintillation light between the SiPMs, caused by the reduction of the light output. 

As shown in Figure~\ref{deltaz.pdf}, the position resolution in the $z$-direction (crystal depth) is not as good as the resolution in the $x$ and $y$-directions. The $z_\mathrm{rec}-z_\mathrm{beam}$ distributions for the 20~mm thick detector have a significant asymmetry, which is caused by the poor $z$ reconstruction of the gamma-ray interactions that occur far from the SiPM array. For such events the reconstructed $z$ is strongly biased towards the centre of the crystal, resulting in large negative values of $z_\mathrm{rec}-z_\mathrm{beam}$.  
At 81~keV, the interaction depth resolution becomes comparable to the thickness of the crystal. In addition, the $z$-reconstruction error at 81~keV has a large negative bias, which means that the ANN trained for $z$-reconstruction with the 356~keV data is not optimal for 81~keV events. This bias is caused by large fluctuations of the relative SiPM signals in the 81~keV sample, which are absent or rare in the 356~keV sample. It should be noted that most 81~keV gamma-rays are absorbed within the first millimetre of CeBr$_3$. Therefore, the calorimeter does not need to measure the interaction depth for such low energy gamma-rays.    

\begin{figure}[!h]
  \begin{center}
    \includegraphics[width=9cm]{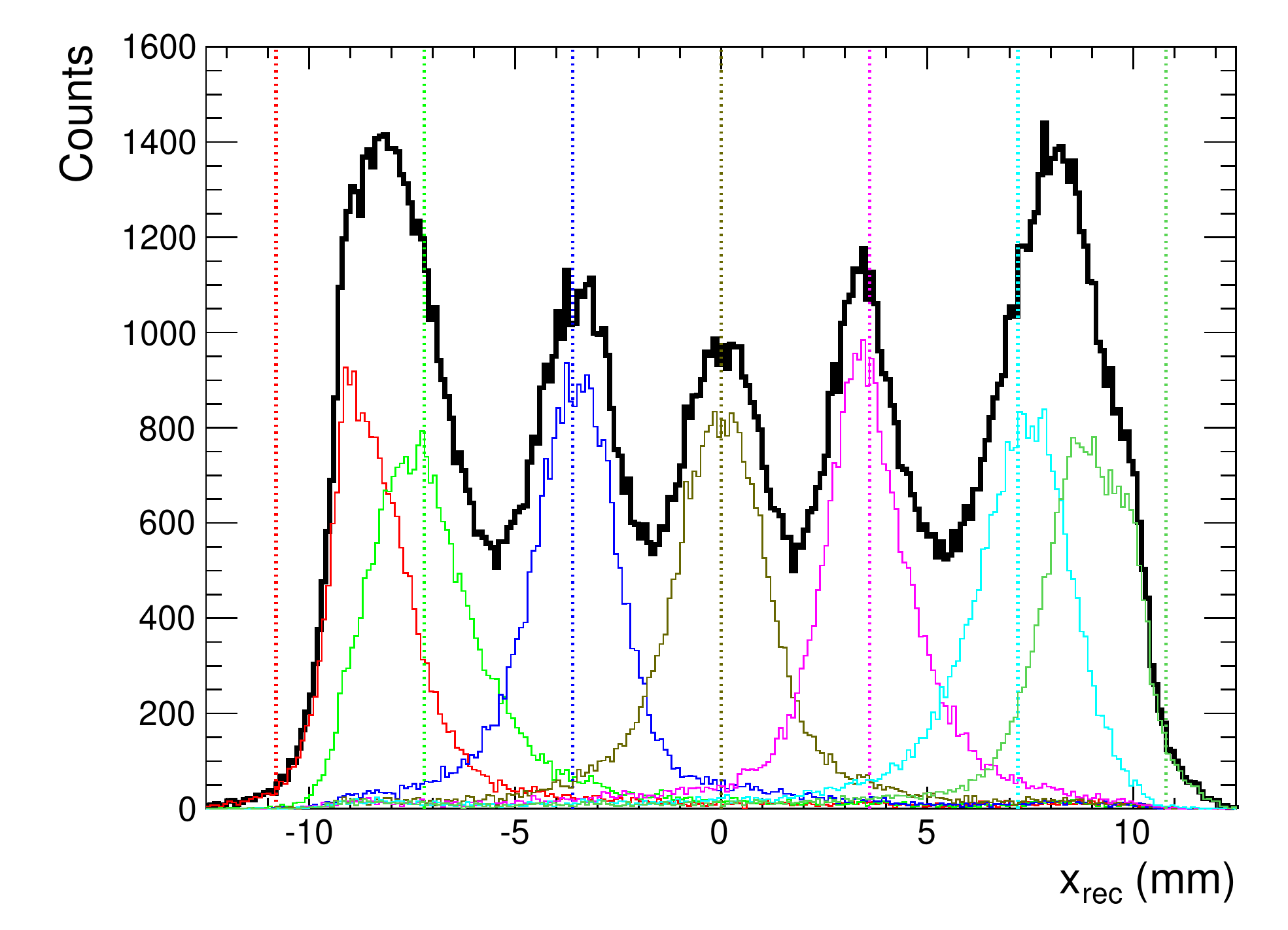}
    \caption{Reconstructed $x$-coordinates for 7 different beam positions along the $x$-axis. The distributions shown by the thin lines correspond to the different beam positions and the thicker line on top shows the total distribution for all beam positions.  The actual $x$-coordinates of the beam are marked by dashed vertical lines. The sides of the crystal correspond to $x=\pm12.5$~mm. This figure is for the 10~mm thick CeBr$_3$ detector and 356~keV gamma-rays.}
    \label{x10.pdf}
  \end{center}
\end{figure}

\begin{figure}[!h]
  \begin{center}
    \includegraphics[width=9cm]{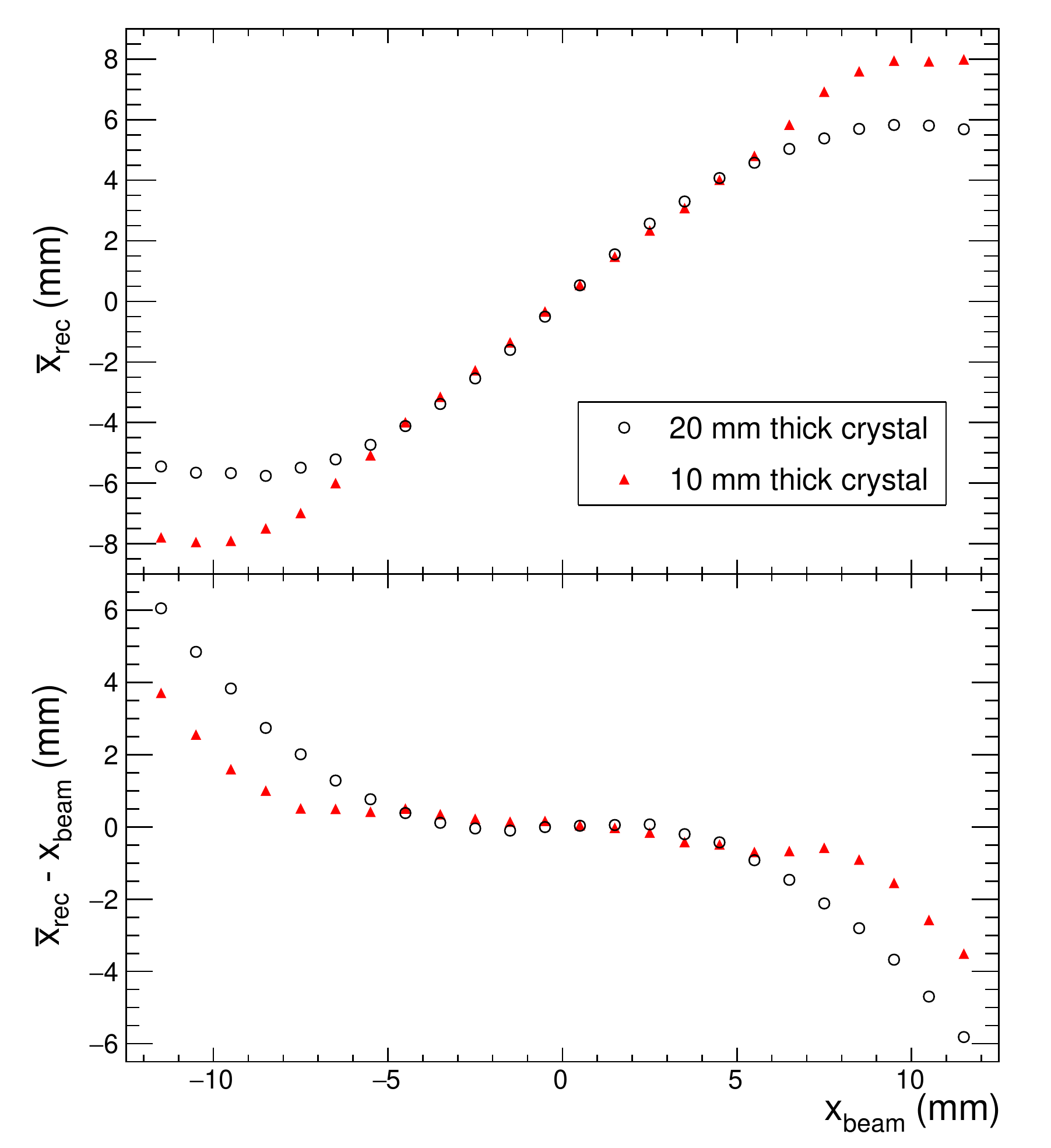}
    \caption{Average reconstructed $x$-coordinate (top) and reconstruction bias (bottom) for 356~keV gamma-rays as a function of the beam position along the $x$-axis. Each data point ($x$-position) uses 24 beam positions along the $y$-axis covering the full size of the crystal.}
    \label{biasx.pdf}
  \end{center}
\end{figure}
\begin{figure}[!h]
  \begin{center}
    \includegraphics[width=9cm]{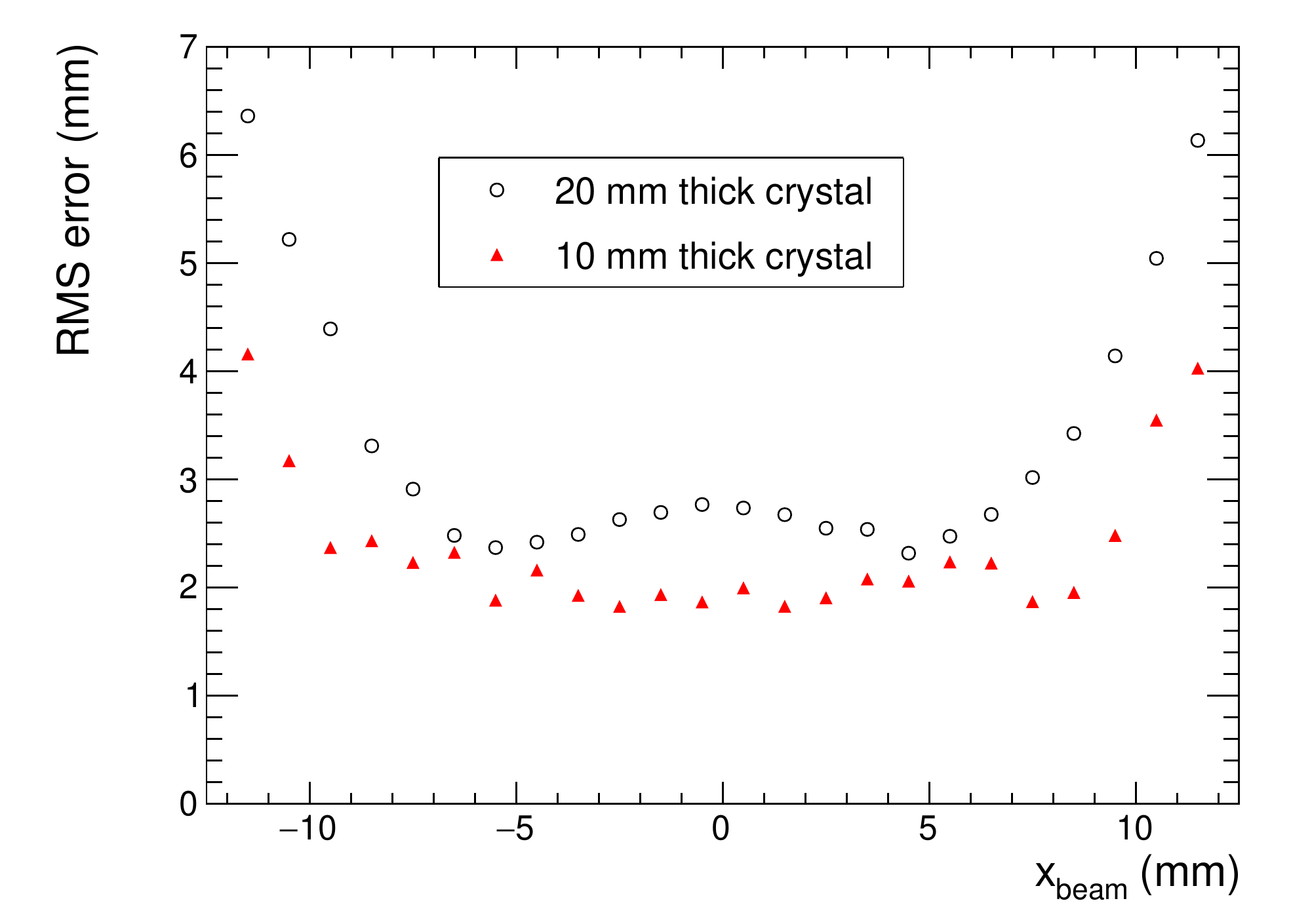}
    \caption{Root-mean-square error of the reconstructed $x$-coordinate for 356~keV gamma-rays as a function of the beam position along the $x$-axis.}
    \label{rmse.pdf}
  \end{center}
\end{figure}

The position resolution of gamma-ray interaction points is not the same throughout the crystal. Reflections from the crystal sides distort the spatial distribution of the scintillation light at the photodetector and make position reconstruction less accurate, particularly if the interaction point is close to a side of the crystal. This effect is illustrated in Figure~\ref{x10.pdf}. The position resolution is best near the centre of the photodetector and gets significantly worse for interactions that occur close to the sides of the crystal. Beam positions separated by $\Delta x=3.6$~mm are easily resolved in the centre of the 10~mm CeBr$_3$ crystal but cannot be resolved near the edges. Figures~\ref{biasx.pdf} and~\ref{rmse.pdf} give more details on how the accuracy of the position reconstruction depends on the gamma-ray interaction position within the crystal. For the 10~mm thick CeBr$_3$ crystal and 356~keV gamma-rays,  the reconstruction bias for the $x$-coordinate is under 1~mm in the central region $|x_\mathrm{beam}| < 9$~mm. Outside this region, the mean reconstructed $x$-coordinate stays nearly constant, which means that the bias increases as the beam approaches the sides of the crystals. The root-mean-square error increases from 2~mm in the central region to over 4~mm when the beam is 1~mm away from the crystal sides. In the case of the 20~mm thick crystal, the reconstruction errors are significantly larger. When the distance $\Delta x$ between the beam and a crystal side is less than 5~mm, the position of the beam has little effect on the reconstructed $x$-coordinate.

A similar study was recently performed using 10~mm thick LaBr$_3$:Ce  and CeBr$_3$ crystals coupled to 64-channel multi-anode PMTs~\cite{gostojic2016}. The crystals were much larger than those used in this work, having a cross-section of about $51 \times 51$~mm$^2$.  For the  LaBr$_3$:Ce crystal, the root-mean-square error of the reconstructed $x$-coordinate was found to be 2.4~mm at 356~keV and 2.9~mm at 59.5~keV.  Those are the average numbers for nine beam positions along a diagonal line of the front face of the crystal, which are very similar to the results obtained for the 10~mm thick CeBr$_3$ crystal in this work: the root-mean-square error for the full XY scan of the front face is 2.4~mm at 356~keV and 2.9~mm at 81~keV. The results obtained in work~\cite{gostojic2016} for CeBr$_3$ were considerably worse, as the ANNs were trained with simulated data and the simulations for the CeBr$_3$ detector did not agree very well with the experimental observations.  

Two other studies reported a position resolution of 1.6~mm FWHM at 511~keV for a $18.0 \times 16.2 \times 10.0$~mm$^3$ LaBr$_3$:Ce crystal~\cite{seifert2012} and 1.4~mm FWHM at 511~keV for a $32.5\times 35 \times 10$~mm$^3$ LaBr$_3$:Ce crystal~\cite{llosa2013}. In both studies the scintillator crystals were coupled directly to SiPM arrays without using light guides. Even after taking into account the difference in the light output of CeBr$_3$ and LaBr$_3$:Ce and the difference in the gamma-ray energy, those studies indicate that the position resolution obtained in this work for the 10~mm thick CeBr$_3$ crystal can be potentially improved by optimisation of the detector, or possibly by using different reconstruction methods.

\subsection{Relative light yield and energy resolution}
The results of the light yield and energy resolution measurements are summarised in Table~\ref{table_eresolution}. In addition to the three crystals used in this study, the PMT measurements were performed for a 51~mm diameter 51~mm thick cylindrical CeBr$_3$ crystal from SCIONIX. This cylindrical crystal has a diffusive interface to the optical window, similar to the LaBr$_3$:Ce crystal. The photoelectron yield of all CeBr$_3$ crystals was found to be about 70\% relative to the LaBr$_3$:Ce crystal, which is consistent with other studies~\cite{quarati_cebr3}. The relative photoelectron yield of the CeBr$_3$
crystals was slightly higher when measured with the SiPM array due to the difference in the spectral sensitivity of the SiPMs and the PMT. 

The energy resolution of 3\% measured for the LaBr$_3$:Ce crystal coupled to the PMT is in line with other studies~\cite{sg, quarati2007}. The energy resolution of the cuboid CeBr$_3$ crystals, however, was found to be slightly worse than the typical value of 4.3\% reported for common cylindrical CeBr$_3$ crystals~\cite{quarati_cebr3}. This difference in resolution can be explained by variations in crystal quality, or can be an effect of the less diffusive interface to the optical window for our cuboid crystals, which makes the distribution of scintillation light over the photocathode less uniform. The energy resolution values obtained with the SiPM 
array are not as good as those achieved with the PMT, which is explained primarily by the lower photon detection efficiency of the SiPMs and the dead space between the SiPMs. It should be noted, however, that 
the development of SiPMs has continued at a fast pace and better SiPM arrays are now available on the market. For example, the new J-series arrays from SensL have an improved photon detection efficiency and minimal dead space. These arrays are therefore expected to provide a considerably better energy resolution compared to the SiPM array used in this work.   

\begin{table}[!h]
\caption{Photoelectron yield relative to LaBr$_3$:Ce and energy resolution (FWHM) for 662~keV gamma-rays.}
\centering
\makebox[\linewidth]{
\begin{tabular}{|l|cc|cc|} \hline
Crystal size (mm) & \multicolumn{2}{|c|}{Photoelectron yield} & \multicolumn{2}{|c|}{Energy resolution} \\
  & SiPM   & PMT   & SiPM & PMT \\
\hline
LaBr$_3$:Ce 28x28x20 & 100\% & 100\% & 4.0\% & 3.0\%\\
CeBr$_3$ 25x25x20 & 70\% &  66\%  & 5.6\% & 4.7\%\\
CeBr$_3$ 25x25x10 & 73\% &  69\%  & 5.4\% & 4.7\%\\
CeBr$_3$ $\diameter 51$x51 & - & 68\% & - & 4.3\% \\
\hline

\end{tabular}
}
\label{table_eresolution}
\end{table}

\begin{figure}[!h]
  \begin{center}
    \includegraphics[width=9cm]{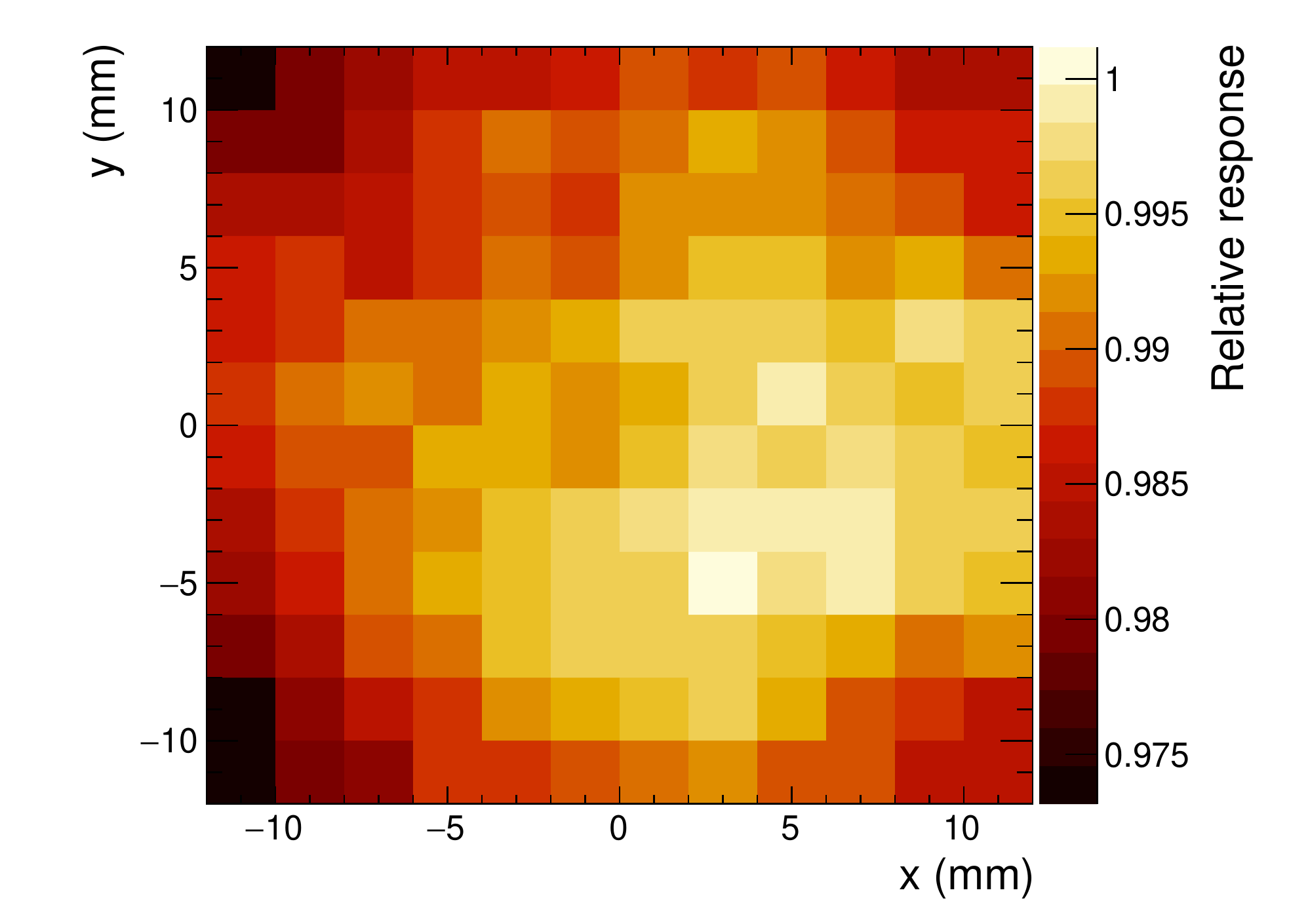}
    \includegraphics[width=9cm]{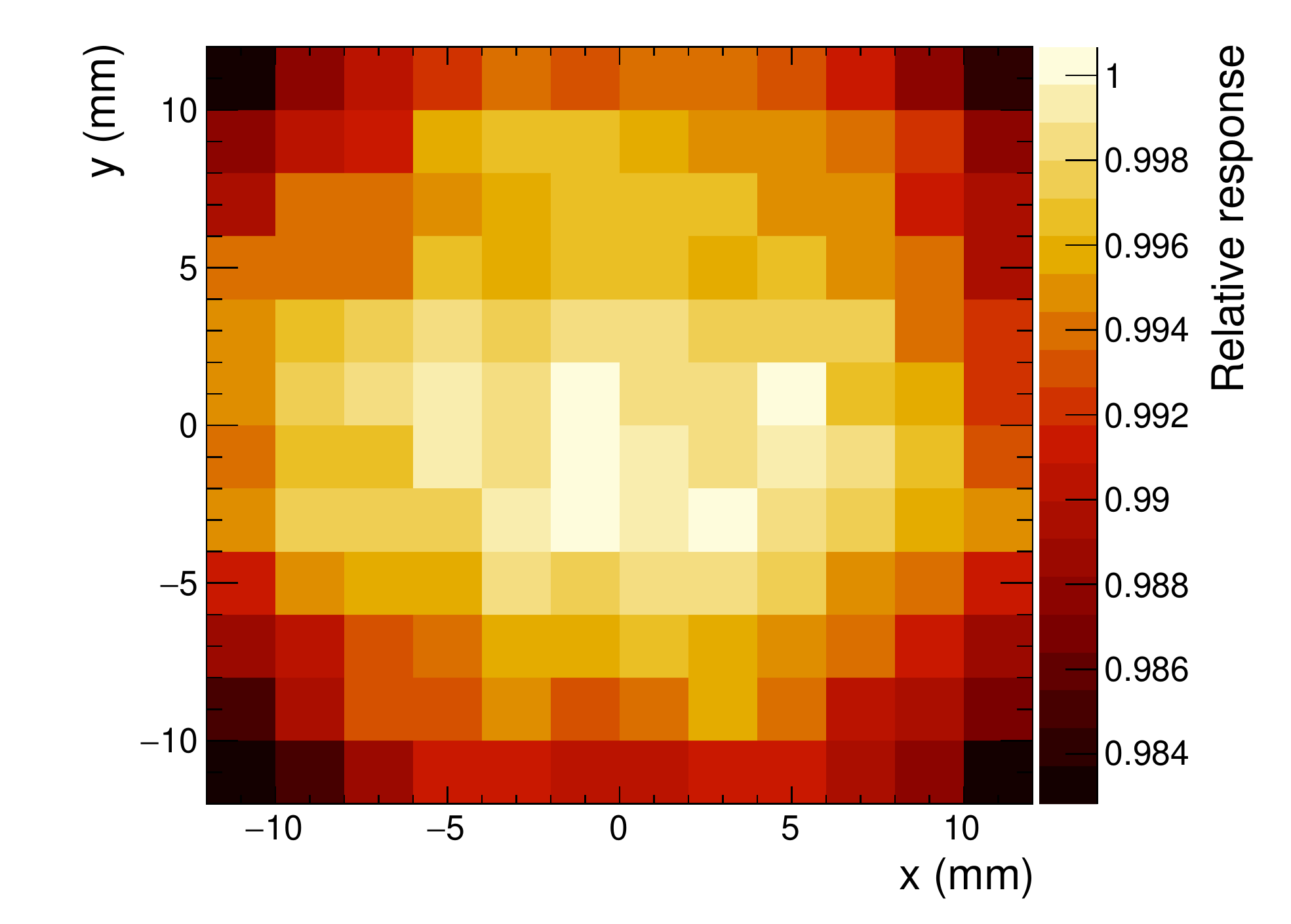}
    \caption{The average detector response to the collimated $^{133}$Ba source calculated in a band of 326-386 keV (full absorption peak for the 356~keV gamma-rays) as a function of the beam position. The top plot is for the 10~mm thick CeBr$_3$ crystal; the bottom plot is for the 20~mm thick CeBr$_3$ crystal.}
    \label{resp10xy.pdf}
  \end{center}
\end{figure}

In Ref.~\cite{gostojic2016}, a multi-anode PMT coupled to an encapsulated $51 \times 51 \times 10$~mm$^3$  LaBr$_3$:Ce or CeBr$_3$ crystal was found to collect about 5\% less light for the gamma-ray interactions that occurred close to the crystal sides than for the interactions in the centre of the crystal. In order to improve the energy resolution of the detector, an energy correction depending on the reconstructed interaction position was used in that study. As shown in Figure~\ref{resp10xy.pdf}, similar response non-uniformity is observed in this work, but the scale of the effect is much smaller, possibly due to the two times smaller X and Y dimensions of the crystals. The maximum signal drop is about 2.5\% in the corners of the 10~mm thick CeBr$_3$ crystal and 1.6\% for the 20~mm thick CeBr$_3$ crystal. The response function for the 10~mm thick crystal is clearly asymmetric, having a maximum at an off-centre position. This asymmetry is an indication of an imperfect crystal or crystal packaging, but may be of little practical importance as the response non-uniformity of the detector is quite small. No correction for the response non-uniformity was used in this work to calculate the energy resolution of the detectors.

\section{Conclusions and perspectives}

This study shows that an encapsulated 20~mm thick CeBr$_3$ crystal coupled to an SiPM array can provide sufficiently good localisation of gamma-ray interaction points inside the crystal, which can be used as the foundation of a calorimeter for a future spaceborne Compton telescope. It should be noted, however, that a thick optical window combined with a diffusive optical interface between the window and the crystal may significantly deteriorate the position resolution of the detector. 
A better position resolution may be potentially obtained by eliminating the optical window and coupling the SiPM array directly to the crystal. For a space mission, the crystal packaging will be optimised to meet the energy and position resolution requirements of the calorimeter. 

In this work, the position resolution was measured for 81~keV and 356~keV gamma-rays and is expected to improve at higher photon energies, as the statistical fluctuations of the SiPM signals get smaller relative to the signal magnitude. There are, however, several factors that limit or may even degrade the position resolution at higher energies, such as multiple photon interactions in the crystal and the substantial range of the electrons and positrons produced by high energy photons (which reaches 12~mm for 10~MeV electrons). Geant4 simulations will be used to study how these phenomena affect the position resolution. 

For efficient detection of high-energy gamma-rays the calorimeter would require several layers of 20~mm thick detectors. Alternatively, a 40~mm thick crystal with a double-sided readout (photodetector arrays coupled to both the front and the back sides of the crystal) is expected to deliver an even better position resolution, as propagation of direct (unreflected) scintillation light to the nearest photodetector is identical to that in the 20~mm thick crystal, while the amount of background (reflected) light is reduced. Further improvements in the position resolution can be expected from increasing the width of the crystals and coupling them to larger SiPM arrays, as this helps to minimise the negative effect of light reflections from the crystal sides.    

The gamma-ray energy resolution of the CeBr$_3$ detector equipped with the SiPM array is not optimal for use in a Compton telescope, but the performance can be improved with development of doped CeBr$_3$ crystals and further advances in SiPM technology, already underway.  

\section*{Acknowledgements}
This work was supported under ESA's Strategic Initiative AO/1-6418/10/NL/Cbi. LH and OJR/SMB acknowledge support from Science Foundation Ireland under grants 11/RFP.1/AST/3188 and 12/IP/1288. DM acknowledges support from Irish Research Council under grant GOIPG/2014/453.

\section*{References}

\bibliography{esa}


\end{document}